\documentclass[preprint,12pt]{elsarticle}

\usepackage{graphicx} 
\usepackage{amsmath}
\usepackage{amssymb}

\usepackage{setspace}
\usepackage[colorlinks=true, allcolors=blue]{hyperref} 
\usepackage[english]{babel}
\usepackage{physics}
\usepackage{tabularx}
\usepackage{amsfonts}
\usepackage{color} 
\usepackage{graphicx, multicol,float} 
\usepackage{caption}
\usepackage{subcaption}
\usepackage{lscape}
\usepackage[english]{babel}	
\usepackage[utf8]{inputenc}
\usepackage{physics}
\usepackage[normalem]{ulem}
\usepackage{orcidlink}
\biboptions{sort&compress}
\usepackage[colorlinks=true, allcolors=blue]{hyperref} 
\usepackage{textcomp}
\emergencystretch=\maxdimen
\newcommand{\REV}[1]{{ \color{blue}  #1}}

\makeatletter
\global\elsprelimpagehlfalse
\makeatother

\begin{document}

\begin{frontmatter}



\title{Mechanical Scaling Laws and Deformation Behavior of Nanoporous Tantalum Microparticles \tnoteref{doi_note}}
\tnotetext[doi_note]{https://doi.org/10.1016/j.jmrt.2026.06.076}

\author[ICB,FCEN,SABATO]{J.I. Ramallo \orcidlink{0000-0002-2353-190X} } 

\affiliation[ICB]{organization={Instituto Interdisciplinario de Ciencias Básicas (ICB), UNCUYO-CONICET-FCEN},
            addressline={Padre Contreras 1300}, 
            city={Mendoza},
            postcode={5500}, 
            state={Mendoza},
            country={Argentina}}

\affiliation[FCEN]{organization={Facultad de Ciencias Exactas y Naturales, Universidad Nacional de Cuyo},
            addressline={Padre Contreras 1300}, 
            city={Mendoza},
            postcode={5500}, 
            state={Mendoza},
            country={Argentina}}

\author[FAMAF,GrupoNICO]{N. Vázquez von Bibow \orcidlink{0000-0003-0551-3561} } 

\affiliation[FAMAF]{organization={Facultad de Matemática, Astronomía, Física y Computación (FAMAF), Universidad Nacional de Córdoba (UNC)},
            addressline={Ciudad Universitaria}, 
            city={Córdoba},
            postcode={5000}, 
            state={Córdoba},
            country={Argentina}}

\affiliation[GrupoNICO]{organization={Grupo de Teoría de la Materia Condensada and
Instituto de Física Enrique Gaviola (IFEG-CONICET)},
            addressline={Ciudad Universitaria}, 
            city={Córdoba},
            postcode={5000}, 
            state={Córdoba},
            country={Argentina}}
            
\author[IMDEA]{M.A. Monclús \orcidlink{0000-0003-3421-3322}} 

\affiliation[IMDEA]{organization={IMDEA Materiales Institute},
            addressline={C. Eric Kandel 2}, 
            city={Getafe},
            postcode={28906}, 
            state={Madrid},
            country={Spain}}

\author[Northwestern]{I. McCue} 

\affiliation[Northwestern]{organization={Department of Materials Science and Engineering},
            addressline={Northwestern University}, 
            city={Evanston},
            postcode={60208}, 
            state={Illinois},
            country={USA}}

\author[INN,SABATO]{M.C. Fuertes \orcidlink{0000-0003-3376-830X}
\corref{cor2}}
\ead{cecilia.fuertes@conicet.gov.ar}

\affiliation[INN]{organization={Gerencia Química and INN, Comisión Nacional de Energía Atómica, CONICET},
            addressline={ Av. Gral. Paz 1499}, 
            city={San Martín},
            postcode={B1650KNA}, 
            state={Buenos Aires},
            country={Argentina}}

\affiliation[SABATO]{organization={ Instituto Sabato, Comisión Nacional de Energía Atómica, Universidad Nacional de General San Martín},
            addressline={ Av. Gral. Paz 1499}, 
            city={San Martín},
            postcode={B1650KNA}, 
            state={Buenos Aires},
            country={Argentina}}

\author[IFN,IMDEA]{C.J. Ruestes \orcidlink{0000-0002-2764-1508} \corref{cor1}}
\ead{cj.ruestes@upm.es}
\affiliation[IFN]{organization={Instituto de Fusión Nuclear "Guillermo Velarde", Universidad Politécnica de Madrid (UPM), ETSI Industriales},
            addressline={C/ José Gutierrez Abascal, 2}, 
            city={Madrid},
            postcode={28006}, 
            state={Madrid},
            country={Spain}}

\begin{abstract}

The mechanical scaling laws of dealloyed nanoporous metals depart from classical Gibson-Ashby predictions for open-cell foams due to a decreased connectivity in their solid network. However, these scaling relations have been established almost exclusively on nanoporous gold produced by electrochemical dealloying, and it is an outstanding question whether the relations apply to nanoporous networks fabricated by other dealloying methods. Here, we investigate the mechanical response of single-crystalline nanoporous tantalum (np-Ta) produced by liquid metal dealloying (LMD) a TiTa alloy in molten CuBi. Nanoindentation of individual microparticles yields an elastic modulus of 10–30 GPa and a hardness of 0.3–1.1 GPa, both scaling with the solid volume fraction in agreement with Gibson-Ashby predictions. This stiffness-density response of np-Ta departs from previous reports on nanoporous gold and is attributed to enhanced ligament connectivity enabled by the thermodynamics of the CuBi metal bath. Molecular dynamics simulations reveal dislocation-dominated plasticity during indentation of np-Ta, consistent with scanning electron microscopy observations of limited densification beneath the indents, ruling out unusual deformation mechanisms as an origin of the observed scaling. These findings identify solvent chemistry in LMD as a tunable lever for ligament connectivity, and thus for the mechanical response of nanoporous metals.

\end{abstract}

\end{frontmatter}



\section{Introduction}
\label{sec:Intro}

Open-cell nanoporous metals, characterized by pores and ligaments at the nanometer scale, are an emerging class of materials with tunable structure and properties \cite{montemayor2015materials,greer2019three}. While relative density has traditionally served as the primary design parameter for cellular solids, advances in synthesis now enable control over the strut morphology, feature size, and hierarchy  \cite{surjadi2025double,surjadi2025enabling,shi2021scaling,vyatskikh2018additive,zhang2021recent}. As a result, these new materials exhibit unique combinations of mechanical and physical properties, which are anticipated to be rapidly adopted across a diversity of applications such as actuation \cite{biener2009surface,detsi2016metallic}, structural \cite{portela2020extreme,shaw2019computationally,wu2023consequences,huber2023densification}, optics and sensing \cite{zhang2013nanoporous}, catalysis \cite{wittstock2023nanoporous}, and nuclear \cite{bringa2012nanoporous,ruestes2025nanoporous}.

Dealloying is a potentially scalable technique to fabricate open-cell nanomaterials, during which the selective dissolution of element(s) from an alloy drives the spontaneous formation of a random porous network. The resulting material can be cubic centimeters in volume, with features as small as 2 nm and solid volume fractions between 0.2 and 0.6 \cite{mccue2016dealloying,mccue2018pattern}. Although the mechanical properties of nanoporous metals have been extensively studied, these reports focused on noble metals, particularly nanoporous gold, whose cost and limited availability constrain use in practical applications. In contrast, more-common base metals (e.g., Fe, Ni, Ti) offer broader technological relevance but are challenging to process via electrochemical dealloying due to their higher reactivity and oxidation tendency  \cite{mccue2018pattern}.

Liquid metal dealloying (LMD) overcomes these processing limitations, replacing aqueous electrolytes and applied potentials with molten metal solvents and enthalpic driving forces. The kinetics and morphological evolution during LMD have been studied over the past decade, enabling the fabrication of a wide range of nanoporous metals, including
Ti \cite{wada2011nano,wada2011dealloying,okulov2017dealloying,berger2020open}, Fe \cite{wada2013three}, FeCr \cite{xiang2020universal,zou2024ligament}, Ta \cite{geslin2015topology,mccue2016kinetics,song2020liquid,lai2022topological}, Nb \cite{kim2014sub,gaskey2019self,sohn2025compressive}, stainless steel \cite{zhao2017three} and even high entropy alloys \cite{okulov2020nanoporous,wada2024accelerated,lee2025development,lee2026nanoporous,choi2026formation}. Despite this progress, mechanical characterization of LMD nanoporous metals remains scarce. Recent studies have shown that the ligament connectivity and morphology of LMD structures are markedly different from those fabricated electrochemically and are linked to both the initial parent alloy composition and the solubility of the remaining component in the liquid metal solvent \cite{lai2022topological}. In turn, these differences call into question the applicability of mechanical scaling laws developed for nanoporous gold, and how the interplay between dislocation activity and densification during deformation may be altered \cite{farkas2018indentation}.

Here, we investigate the mechanical behavior of the prototypical LMD system, nanoporous tantalum (np-Ta), to determine whether scaling laws developed for electrochemically dealloyed nanoporous gold remain predictive. Beyond its potential use in high-power electronics and biomedicine \cite{chakraborti2016ultrathin,wauthle2015additively}, np-Ta is of interest for extreme environments, where refractory metals are needed as heat-pipe wicks and plasma-facing components \cite{peters2024materials,cunningham2024alloying, garrison2023review}.
To isolate the intrinsic foam response from grain-boundary effects, single-crystalline (single-grain-derived) np-Ta microparticles were experimentally fabricated  \cite{chuang2022powder}  and tested via nanoindentation. The resulting mechanical response was then analyzed within established scaling frameworks for cellular solids and compared against previous studies on nanoporous gold and other LMD-derived nanoporous metals. Complementary atomistic simulations were also employed to rule out unusual deformation mechanisms as the origin of the observed trends. The resulting picture points to ligament connectivity, tuned by solvent chemistry during LMD, as the dominant factor setting the stiffness of np-Ta.

\section{Materials and Methods}
\label{sec:Methods}

\subsection{Sample preparation}
\label{sec:SamplePrep}

Ti$_{65}$Ta$_{35}$ alloys were prepared in-house by radio-frequency (RF) induction using an Ambrel Ekoheat 45 kW system by melting Ti (99.995 wt.\%), and Ta (99.95 wt.\%) pellets from Kurt Lesker in a water-cooled copper crucible from Arcast Inc. under flowing Ar (99.999 wt.\%). After casting, the ingots were annealed under flowing Ar (99.999 wt.\%) for approximately 10 h. Compositional homogeneity was confirmed using energy-dispersive X-ray spectroscopy (EDS). Master ingots of $\sim$30 g were made, rolled to foils $\sim$100 $\mu$m thick, and then annealed at 1200 °C for 8 h. 

\begin{figure}[H]
\centering
\includegraphics[width=1.0\textwidth]{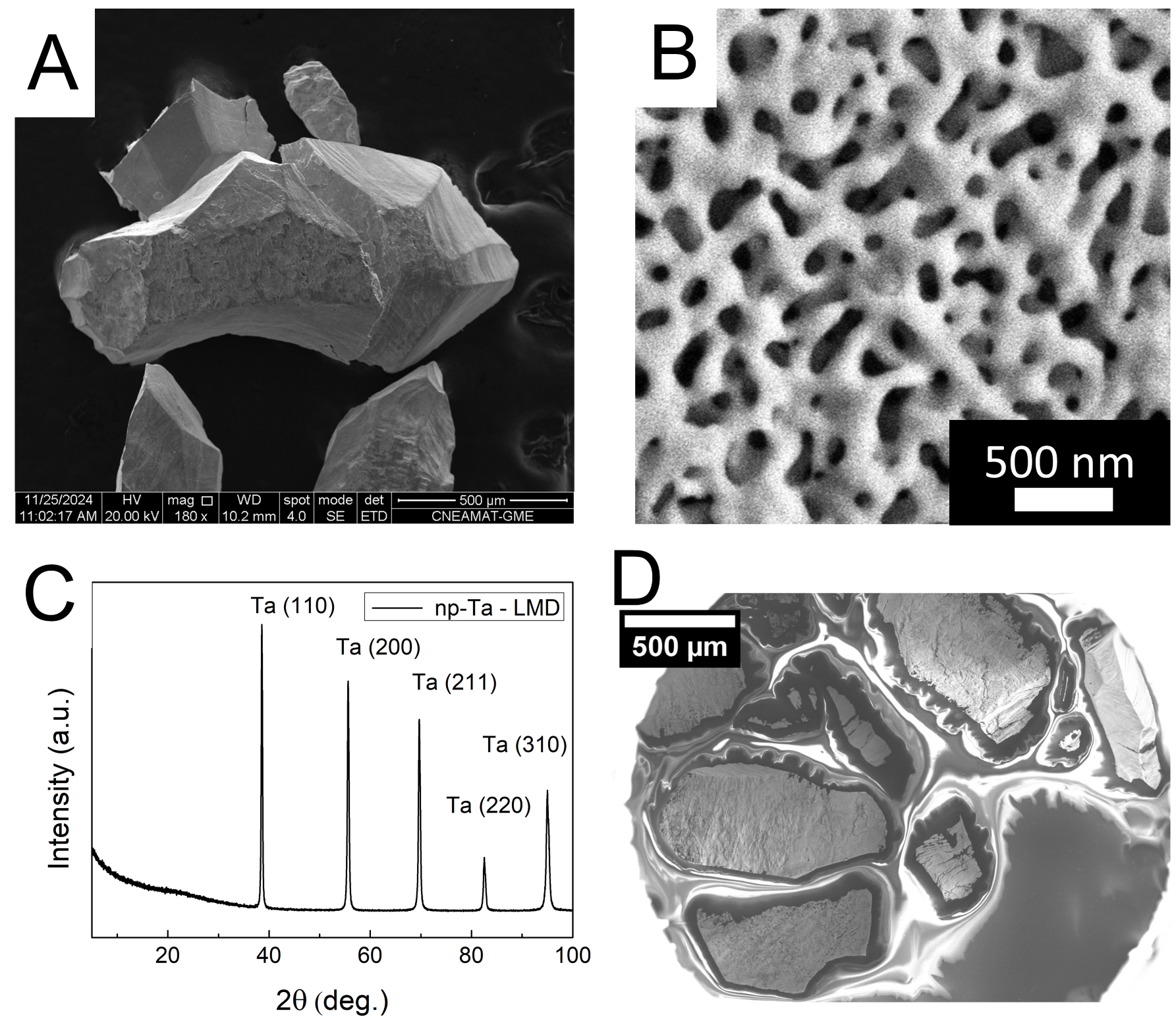}
\caption{A. SEM image of typical np-Ta particles. B. SEM image of the surface showing the nanoporous bicontinuous structure of one particle. C. XRD pattern of the np-Ta powder sample. D. Particles deposited on top of a Si wafer  using CrystalBond \textsuperscript{\textregistered} 509 (bright phase bounding particles) for nanoindentation testing.}\label{fig:Sample}
\end{figure}

Dealloying was carried out by immersing foils into a 40 g bath of molten Cu$_{40}$Bi$_{60}$ at $\sim$800 °C in an ultra-high purity alumina crucible cast in-house using materials from Cotronics Corp. Induction melting was used to heat the samples under flowing Ar (99.99\%), and the LMD process was run for 5 min. After dealloying, the samples were immersed in concentrated nitric acid to remove the CuBi and then rinsed with de-ionized water. The resulting material consisted of nanoporous powders (Fig. \ref{fig:Sample}.A,B) and the ligament diameter and solid volume fraction ($\phi$) were quantified using AQUAMI \cite{stuckner2017aquami,mccue2018gaining}. The solid fraction is estimated from surface scanning electron microscopy and thus is an approximate 2D proxy. X-ray diffraction (XRD) measurements were performed to determine the presence of crystalline phases of tantalum (Fig. \ref{fig:Sample}.C). A PANalytical Empyrean X-ray diffractometer (INN-CNEA-CONICET) equipped with Cu K radiation ($\lambda$ = 1.54 \AA) and a PIXcel3D area detector was used. \REV{The powders were placed on a glass sample holder to avoid signals from the support and measured without additional preparation.} Morphology and composition of particles obtained were evaluated using scanning electron microscopy (SEM). Investigations were performed using a FEI Quanta 200 microscope equipped with EDS (EDAX$ \textsuperscript{\textregistered}$)(Gerencia de Materiales, CAC-CNEA) and a FEI Helios NanoLab 600i FEGSEM microscope (IMDEA Materials Institute). \REV{For the EDS measurements, conductive carbon tape was placed on the SEM stub, and a small amount of powder was deposited onto the adhesive surface using a spatula tip. Five particles were analyzed, and three spectra were acquired from each particle (see examples in Fig. S2, in the supplementary material file).}

\subsection{Nanoindentation}
\label{sec:NanoMeth}

To carry out the nanoindentation experiments,
np-Ta powder particles were immobilized on polished Si substrates for nanoindentation measurements (Fig, \ref{fig:Sample}.D) by heating the Si substrate to $\sim$70 °C, applying CrystalBond$ \textsuperscript{\textregistered}$ 509 forming an homogeneous thin film on its surface, and then depositing the particles directly using a spatula. Upon cooling, the resin hardens, effectively bonding the particles to the substrate.

Hardness (H) and indentation elastic modulus (E) were then obtained using nanoindentation. Measurements were performed on two independent platforms (Agilent G200-XP and  Hysitron TI950), both with a Berkovich diamond tip, to verify reproducibility. \REV{The area function of the tip was calibrated by performing a series of indentations with increasing load on a standard fused silica sample.}

Depth-dependent averages for the elastic modulus were obtained via the continuous stiffness method 
 (nanoDMA-Hysitron Triboindenter) with a constant strain rate of 0.1 $s^{-1}$ and load-controlled mode up to a load of 100 mN. Load oscillation was adjusted to 75 Hz using variable load amplitude (starting at 175 $\mu$N) to obtain a harmonic displacement target of 2 nm. Load-displacement curves were measured up to 1000 nm using both indenters with displacement-controlled mode,\REV{ reaching maximum loads between 12 and 25 mN.  The selected indentation depth minimizes substrate effects, as it is smaller than 10\% of the average particle size.} 
 
 \REV{Nanoindentation measurements were carried out on 9 different particles, all of which exhibited flat exposed surfaces. Optical images of the indented particles are presented in Fig. S1, in the supplementary material file.} Arrays of \REV{5 to 15 imprints per particle, depending on the particle size,} were performed with a minimum distance between indents of 50 $\mu$m to prevent interference from deformation caused by nearby indents \cite{baker2024optimal}. 
A strain rate of 0.05 1/s and a holding time at maximum load of 10 s were used. As a first approximation, a Poisson's ratio of 0.18 was utilized for \REV{modulus} calculations \cite{luhrs2016elastic}.

\subsection{Simulation methods}
\label{sec:SimMeth}

Molecular dynamics nanoindentation simulations were employed \cite{ruestes2017molecular,janisch2025nanoindentation} considering a np-Ta substrate and a Berkovich tip. The np-Ta substrate was generated using the levelled-wave method \cite{soyarslan20183d} and has been previously characterized in \cite{von2024topological}. Further details are available in the supplementary materials file. The Berkovich tip model is a three-face pyramid whose corner was rounded with a 10 nm fillet radius, tangent to the pyramid edges. Unlike previous studies using virtual indenters, a tip with atomistic representation was used, following the methods presented in earlier work \cite{ruestes2017molecular,zonana2020effect,alhafez2021indentation}. The atomistic tip is composed of carbon atoms arranged in a rigid diamond lattice structure. The indentation axis is centered with respect to the substrate. In order to avoid tip-substrate interactions prior to indentation, the initial position of the indenter is 1 nm above the substrate.  

The indenter tip penetration rate was chosen as 3.4 m/s. This velocity, though high compared to experimental indentation, is well below the longitudinal wave velocity of Ta ($\approx$ 3400 m/s) and should be considered low enough for typical MD studies. Ruestes et al. \cite{ruestes2014atomistic} and Alcalá et al. \cite{alcala2012planar} performed MD simulations of bcc Ta nanoindentation for velocities of 3.4–34 m/s and 4–0.004 m/s, respectively; they reported that the plasticity mechanisms showed no significant differences in this velocity range. Variations in the indentation velocity had a low influence in terms of the extension of the plastic zone  \cite{ruestes2014atomistic}.
The MD simulations were carried out using the open-source LAMMPS code \cite{thompson2022lammps} with a constant time step of 1 fs. Visualization was performed using OVITO \cite{stukowski2009visualization} and dislocations were identified using DXA \cite{stukowski2010extracting}.

\section{Results and Discussion}
\label{sec:results}

\subsection{Material Microstructure and Initial Mechanical Response}

Fig. \ref{fig:Sample}.A presents a low-magnification SEM image of the np-Ta microparticles obtained after liquid metal dealloying. The particles exhibit a non-spherical and polyhedral shapes, arising from detachment of individual grains from the parent foil during dealloying due to preferential grain-boundary corrosion in the molten Cu–Bi bath \cite{mccue2016kinetics,bieberdorf2023grain,kerr2024morphology}. 
As a result, each particle is a single grain inherited from the Ti$_{65}$Ta$_{35}$ precursor, and examination at higher magnification (Fig. \ref{fig:Sample}.B) reveals the open-cell network of Ta ligaments with an average ligament diameter of 200 $\pm$ 100 nm and a solid volume fraction of $\phi \approx 0.35$. These values are in good agreement with prior reports using a similar protocol \cite{chuang2022powder}. 
\REV{The initial nanoporous length scale is set by surface roughening at the dissolution front \cite{erlebacher2003pattern}, but at the high homologous temperatures of LMD the structure coarsens rapidly behind the front (following a $t^{0.25}$ power-law behavior) \cite{mccue2016kinetics}. Unless specimens are exposed to the molten metal bath beyond $\sim$60 minutes, temperature is the
dominant control parameter, and the ligament diameter values reported here are representative of the Ti–Ta / Cu–Bi system \cite{song2020liquid,wu2025formation}. The solid volume fraction and ligament morphology are set largely by the parent-alloy composition, with higher Ta content in the precursor yielding higher $\phi$ and more interconnected networks. Within the accessible ligament-size range in this study, no systematic dependence of $E/E_{bulk}$ on
ligament size is observed, so the scaling and connectivity arguments developed below apply.}

The X-ray diffraction pattern of the dealloyed powder (Fig. \ref{fig:Sample}.C) shows only bcc Ta, confirming phase purity and the complete removal of Ti, Cu, and Bi from the material. No crystalline Ta-based oxide phases were detected within the sensitivity of the measurement. EDS of the particles (Fig. S2) returns a composition of Ta with a small oxygen signal (<2 wt.\%), though SEM/EDS has limited quantitative sensitivity to light elements and this value should be interpreted as an upper bound on oxide contamination rather than a direct measurement.

\REV{Nanoporous Ta combines a base material prone to oxidation with a high specific surface area, such that the formation of a native tantalum oxide layer during sample transfer and handling cannot be excluded. Such an oxide may locally affect the mechanical response of ligament surfaces, particularly hardness (e.g. $Ta_2 O_5$ hardness is 16 GPa, while bulk metallic Ta is around 8–10 GPa \cite{melnikova2021nanomechanical}). However, native oxide films formed on Ta under ambient conditions are typically only a few nanometers thick, generally below 10 nm \cite{khanuja2009xps}, whereas the ligament diameter measured here is approximately 200 nm. In addition, no crystalline Ta-based oxide phases were detected by XRD (Fig. \ref{fig:Sample}.C), and SEM/EDS (Suppl. Fig. S2) showed only a small oxygen signal below 2 wt.\% in the Ta powder, although this value should be interpreted cautiously because of the limited sensitivity of SEM/EDS to light elements. 
Altogether, these observations indicate that while a thin amorphous native oxide layer may be present, its thickness is expected to be extremely thin compared to the average ligament size. Therefore, it is not expected to significantly influence the foam-level modulus and hardness measured at 1000 nm indentation depth, where the contact area averages the response over many ligaments.}

\begin{figure}[H]
\centering
\includegraphics[width=0.5\textwidth]{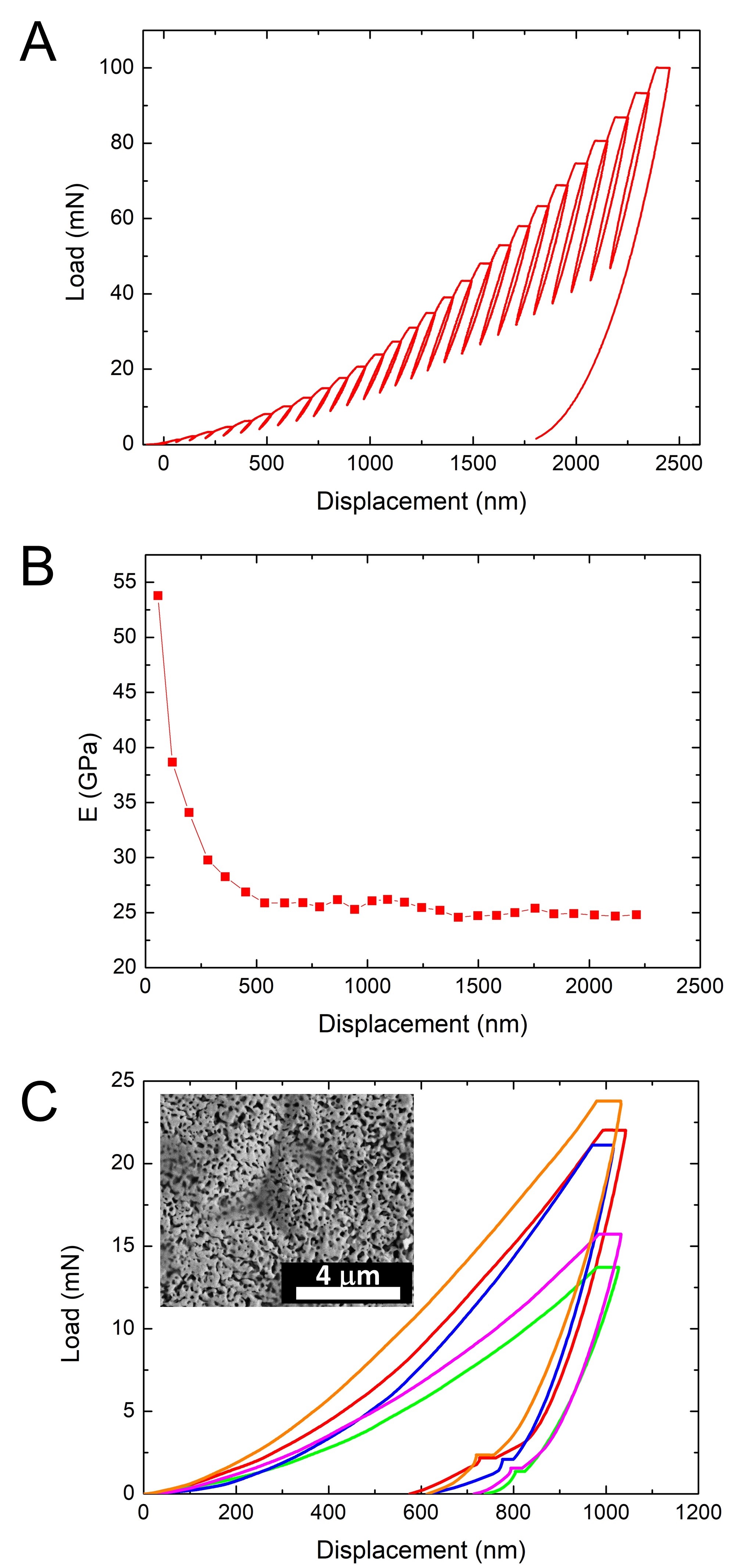}
\caption{A. Continuous stiffness measurement (CSM) load–displacement  curve. B. Depth-dependent averages for the elastic modulus E under CSM. C. Representative nanoindentation curves  up to 1000 nm depth. Color coding corresponds to different np-Ta particles. \textit{Inset}- Top-view SEM image of a residual nanoindentation imprint.
}\label{fig:NI_curves}
\end{figure}

Nanoindentation was carried out on the immobilized np-Ta particles, as described in the Methods section, and only particles exhibiting a clearly defined flat plane parallel to the holder surface were indented. 
Fig. \ref{fig:NI_curves}.A shows a representative CSM load - displacement curve obtained indenting an individual particle. The elastic modulus decreases with increasing indentation depth and reaches a plateau at depths greater than 500 nm, Fig. \ref{fig:NI_curves}.B, consistent with previous reports on nanoporous gold \cite{hodge2009ag,huber2023densification}. Following the suggestion by Hodge et al. \cite{hodge2009ag}, a target depth of 1000 nm (within the plateau region) was selected, and systematic indentations were performed in displacement-controlled mode. At this depth, the estimated contact area $A_c \approx 24.5 \mu m^{2}$ corresponds to a circular contact of radius $r = \sqrt{(A_c/\pi)}=2.8~ \mu$m, which, for a ligament size $l \approx200$ nm, gives $r/l \approx 14$. This ratio is  considered adequate for treating the indented volume as representative of the bulk foam \cite{lilleodden2018topological,soyarslan20183d}, and the measured E and H can therefore be interpreted as homogenized foam-level properties. For the remainder of this article, the reported values of E and H correspond to displacement-controlled indentations performed to a target depth of 1000 nm. 

Fig. \ref{fig:NI_curves}.C presents representative load–displacement curves obtained on several particles, where the plastic behavior of the porous material is evidenced by the hysteresis in the load-displacement response \cite{fischer2004nanoindentation}. The standard Oliver-Pharr method was used to obtain E and H from load-displacement curves \cite{oliver1992improved}. This method has been extensively used in previous studies on nanoporous metals \cite{hodge2007scaling,hodge2009ag,biener2011ald,kreuzeder2015fabrication,leitner2016interface,esque2016nanomechanical,kang2018microstructural,kim2018indentation,joo2020beating,zhao2020open,siddique2023diamond,huber2023densification,baker2024optimal}. Post-indentation SEM inspections (inset of Fig. \ref{fig:NI_curves}.C) reveal limited densification beneath the indenter and no visible cracking at the tested depths, supporting the applicability of the Oliver–Pharr analysis. Progressive densification or ligament collapse during unloading, which could require densification-aware approaches \cite{kwon2021compressive,champion2019understanding,cheng2023investigation,huber2023densification}, appears to be negligible under the present conditions. In addition, the load–displacement curves show a  smooth response during both loading and unloading, with no detectable pop-in or pop-out events, indicating the absence of measurable crack propagation in the indented region. Reproducibility was further confirmed by measurements performed on  two independent nanoindentation platforms, as described in section \ref{sec:NanoMeth}. Variations in slope and maximum load are nevertheless observed for the same target displacement, likely arising from local heterogeneities affecting E and H. 

\begin{figure}[H]
\centering
\includegraphics[width=0.85\textwidth]{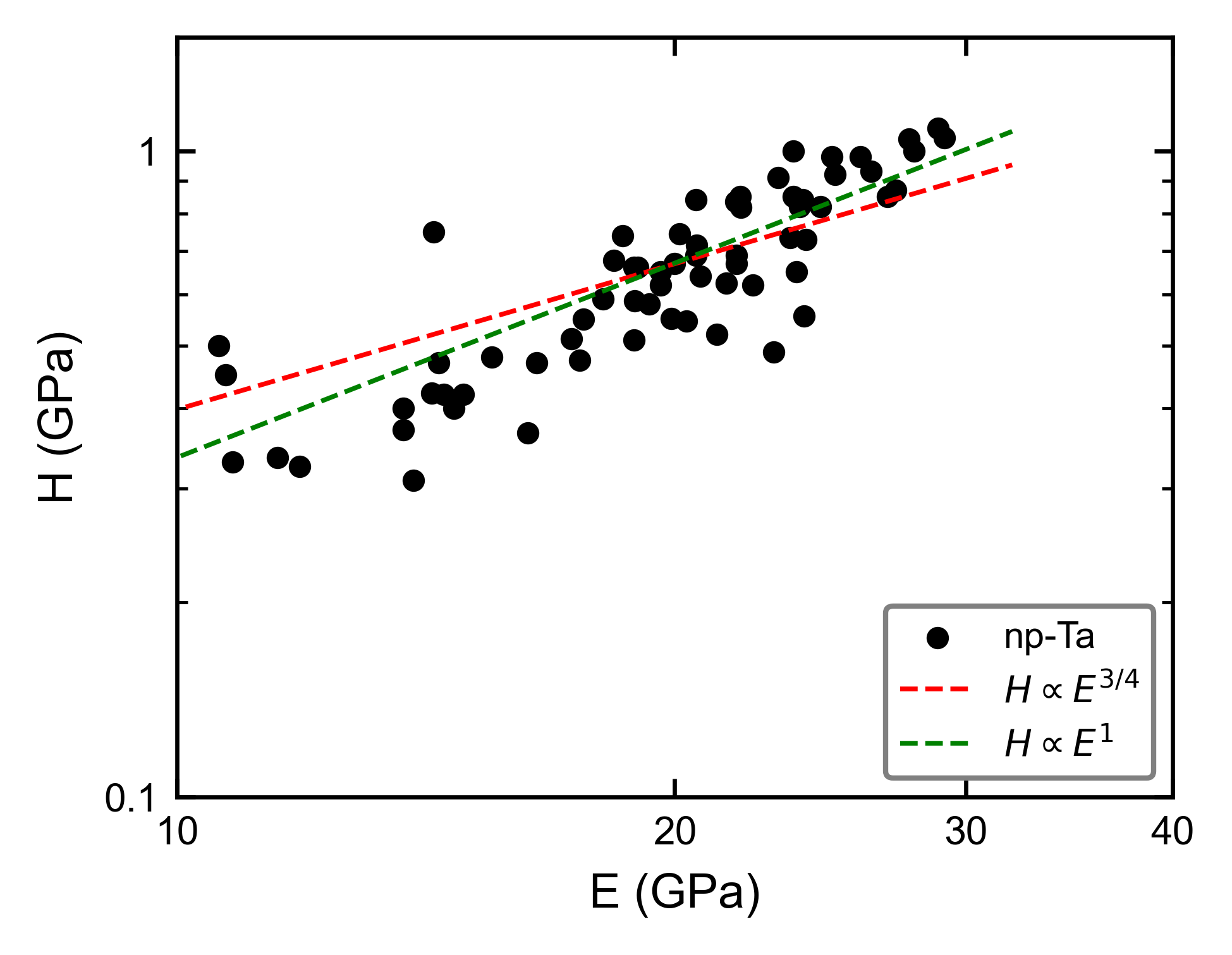}
\caption{Hardness (H) vs elastic modulus (E) values compared with a power law dependence based on equation \ref{eq-H-E}. The experimental uncertainty in E and H is approximately 10\%.
Red line corresponds to $n=3/2$ while the green line corresponds to $n=2$.
}\label{fig:NI_summary}
\end{figure}

The nanoindentation measurements across the population of indented particles are presented as a scatterplot in Fig. \ref{fig:NI_summary}, 
with E values in the range 10–30 GPa and H values in the range 0.3–1.1 GPa. \REV{This plot includes the results obtained from 67 valid measurements performed on the different particles studied.} To test whether this scatter reflects genuine specimen-to-specimen variation in foam architecture rather than measurement noise, the data is analyzed based on Gibson–Ashby scaling laws \cite{gibson1982mechanics}. These equations relate the elastic modulus (E) and yield stress ($\sigma$) of the foam with their respective ligament material properties  ($E_{lig},\sigma_{lig}$). The foam properties are then related to those of the ligaments.

Briefly, E scales as

\begin{equation}
    E = C_E ~ E_{lig} ~ \phi ^2
    \label{eq-GA_E}
\end{equation}

with $C_E$ typically taken equal to 1 and $\sigma$ scales

\begin{equation}
    \sigma = C_\sigma\ ~ \sigma_{lig} ~ \phi ^ {n}
    \label{eq-GA_S}
\end{equation}

with $1 \leq n \leq 2$, typically taken as $n=3/2$ \cite{ashby2000metal}. 

Considering a yield strength-to-hardness ratio $f$ \cite{fischer2004nanoindentation}, 

\begin{equation}
    H = f~\sigma = f~ C_\sigma\ ~ \sigma_{lig} ~ \phi ^ {n}
    \label{eq-H-S}
\end{equation}

Rearranging eqns. \ref{eq-GA_E} through \ref{eq-H-S} renders

\begin{equation}
    H = \frac{f~C_{\sigma}~\sigma_{lig}}{{E_{lig}}^{\frac{n}{2}}}~E^{\frac{n}{2}} = C'~E^{\frac{n}{2}}
    \label{eq-H-E}
\end{equation}

It is worth noting that to account for the effect of network connectivity, the community has proposed a modification introducing the concept of effective solid volume fraction, which is the solid volume fraction of the load bearing network only \cite{liu2016interpreting,liu2017scaling}, as well as modified scaling laws to consider quantitative measures of connectivity, such as the genus \cite{sohn2024scaling}. These considerations provide useful context for interpreting the scaling behavior discussed below.

Fig. \ref{fig:NI_summary} includes equation \ref{eq-H-E} using two sets of parameters. First, a standard Gibson-Ashby set of parameters (i.e. $C_E=1$, $C_{\sigma}=0.3$ \cite{liu2016interpreting}, $n=3/2$ \cite{ashby2000metal} together with $f=2.7$ \cite{ashby2000metal}, $E_{lig}=E_{bulk}=186~GPa$ \cite{ruestes2014atomistic} and $\sigma_{lig}=4.4~GPa$, taken as the yield stress of a nanowire of the sample material \cite{liu2016interpreting}). Second, considering $C_{\sigma}=0.5$ and $n=2$, different from standard parameter sets but within model bounds \cite{ashby2000metal}. These two sets of parameters render $H \propto E^{3/4}$ and $H \propto E^{1}$, respectively. The latter displays excellent agreement with our experimental results, with the non-standard exponent in the solid-fraction power-law dependence, likely reflecting differences in architecture between the open-cell foams underlying the Gibson–Ashby model and the geometry of np-Ta \cite{mccue2016dealloying}.
Overall, the measurements fall within a region that, to a first approximation, is consistent with the relation given in Eqn. \ref{eq-H-E}. Based on the functional form of the underlying equations (Eqns. \ref{eq-GA_E} and \ref{eq-GA_S}), the observed distribution in the data suggests that this behavior arises from variations in the solid volume fraction. 

\begin{figure}[H]
\centering
\includegraphics[width=0.7\textwidth]{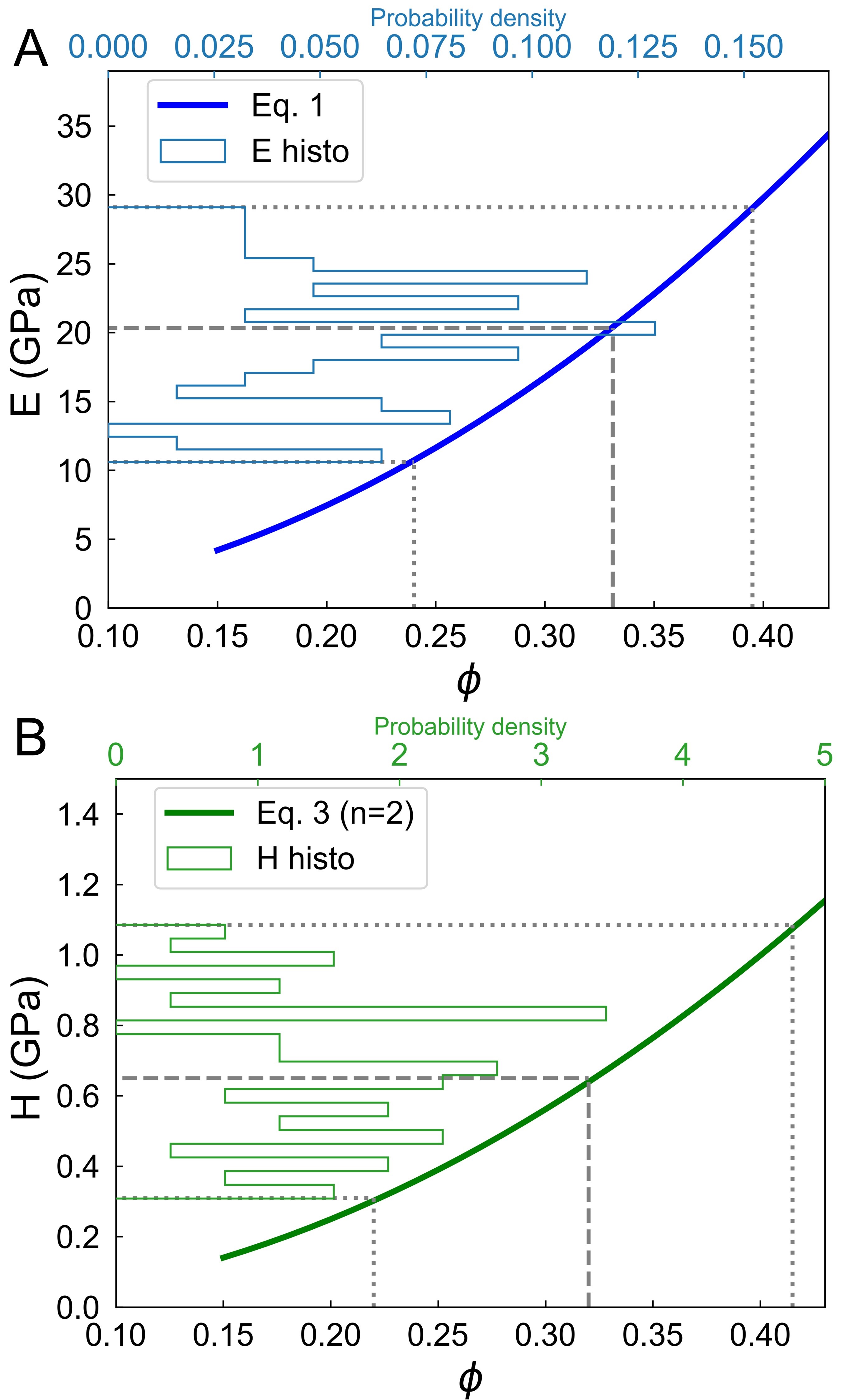}
\caption{Scaling laws predictions for E (Eq. \ref{eq-GA_E}) and H (\ref{eq-H-S}) compared with histograms of nanoindentation results. Dashed lines correspond to the median of values, while dotted lines correspond to limiting values. Note that the median of histogram data closely aligns with the measured solid fraction.}\label{fig:Histo} 
\end{figure}

To further clarify this aspect, Figure \ref{fig:Histo} shows probability density histograms for E and H, compared to the corresponding scaling equations (Eqns \ref{eq-GA_E} and \ref{eq-H-S}, respectively). Comparison of the corresponding scaling equation with the histogram data provides additional indications that the observed scattering in E and H results are due to solid fraction effects. 

The prediction of eq. \ref{eq-GA_E} for $\phi$ in the range  0.24 - 0.39 is consistent with the histogram of E values (Fig. \ref{fig:Histo}.A). Similarly, the prediction of eq. \ref{eq-H-S} with $n=2$ for $\phi$ in the range 0.22 - 0.42 agrees with the histogram of H values (Fig. \ref{fig:Histo}.B). Notably, the mechanically inferred ranges of apparent $\phi$ overlap the image-based SEM/AQUAMI estimate ($\phi = 0.35$), supporting the interpretation that specimen-to-specimen variations in solid fraction are a primary contributor to the scatter in E and H. The differences between E and H-derived solid volume fractions are consistent with the deviations observed in Fig. \ref{fig:NI_curves}. The upper and lower bounds of the measured data correspond to variations in solid volume fraction comparable to those reported previously for np-Ta \cite{mccue2016local}. The scatter observed in the present results can be largely attributed to variations in solid fraction.

\REV{The Gibson–Ashby relations are used here as a reference scaling framework rather than as a material-specific constitutive law for bcc np-Ta. Their functional dependence on relative density reflects foam architecture and load transfer, and is not explicitly tied to a particular crystal structure or slip-system geometry. Crystal structure enters through the ligament-level properties, such as elastic modulus, yield strength, and plastic deformation mechanisms. Accordingly, the present analysis uses Ta-specific ligament properties where possible. Nevertheless, the numerical prefactors and effective exponents may vary between fcc np-Au and bcc np-Ta because they are sensitive to connectivity, ligament morphology, surface state, and ligament-level plasticity. Thus, the agreement with Gibson–Ashby-type scaling should be interpreted as a first-order comparison, while a fully bcc-specific scaling law for nanoporous refractory metals would require systematic measurements over a broader range of densities and topologies.}

\subsection{Insights in Nanoscale Deformation Mechanisms and New Scaling Laws}

The preceding analysis attributes the scatter in E and H to variations in solid volume fraction, but it does not address whether these differences (between the present study and previous reports) arise from additional deformation mechanisms that may be active in np-Ta, especially considering that most prior studies were conducted on np-Au which has a different crystalline structure. Previous MD work by two of this manuscript's authors on highly strained np-Ta \cite{von2024topological}
provides evidence of twinning, which could modify the expected scaling behavior. Although direct observation of defects in the deformed ligaments beneath the indents by transmission electron microscopy would be valuable, extreme care must be taken during focused ion beam lamella preparation to avoid introducing damage to these nanoscale features. In addition, the ligament diameters here are comparable to typical lamella thicknesses and thus can easily be broken during lift-out. As an alternative route, MD simulations of nanoindentation were employed here as an in-silico computational microscopy technique \cite{zepeda2021atomistic} to provide fundamental understanding on the deformation behavior of np-metals (rather than absolute indentation loads or moduli) \cite{xia2015role,ngo2015anomalous,ruestes2016hardening,farkas2018indentation,shi2021scaling,valencia2022nanoindentation,engelman2025deformation,przybilla2025revealing}.

\begin{figure}[h!] 
\centering
\includegraphics[width=0.95\textwidth]{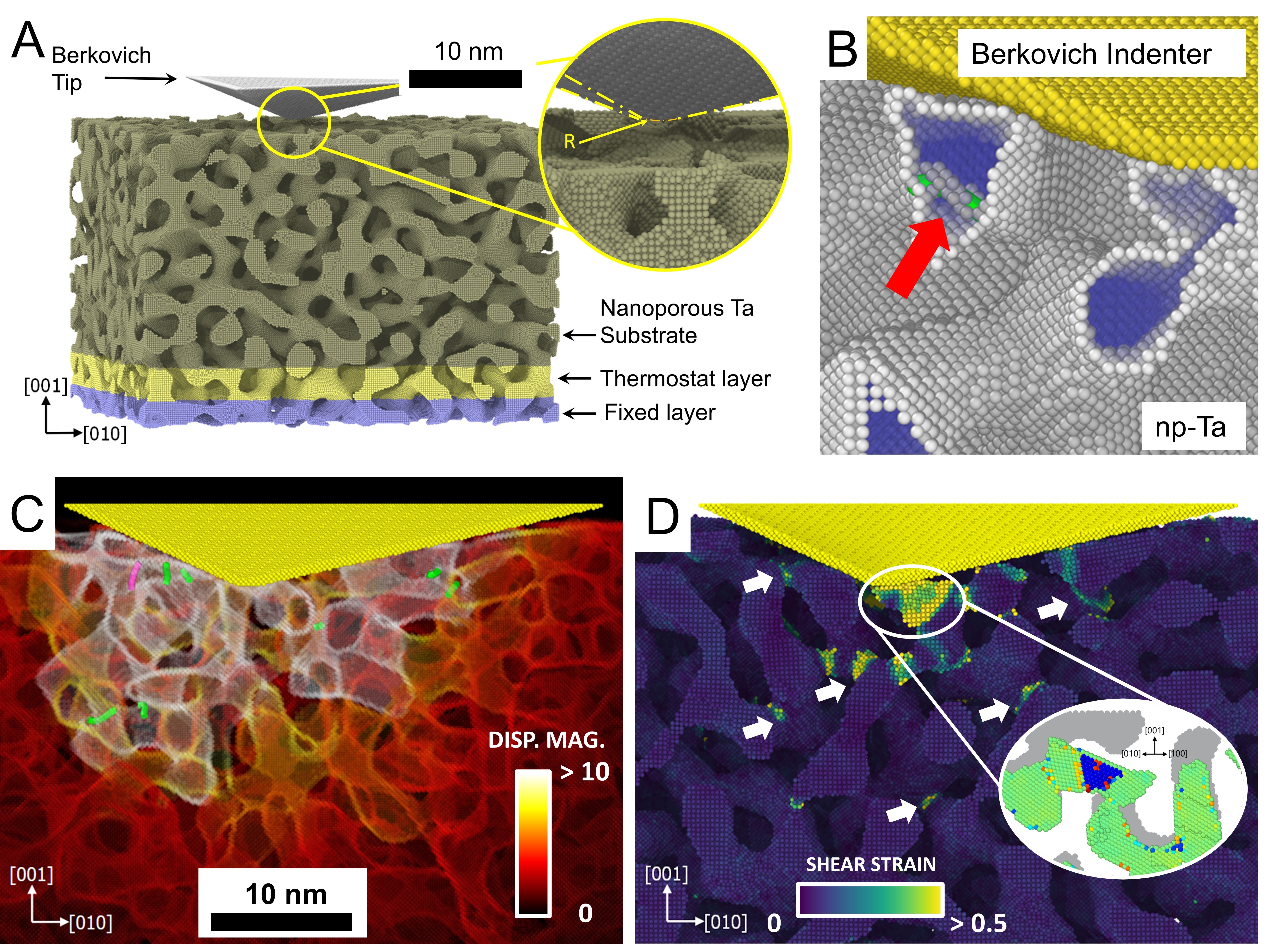}
\caption{A.  Setup of the simulated np-Ta system. B. The penetration of the indenter induces the nucleation of dislocations on the ligament surfaces and their propagation inside the ligaments (red arrow). C. Cross-section indicating the displacement magnitude profile of the ligaments at a penetration depth of 6 nm, together with glissile (green lines) and sessile (pink) dislocations. Transparency was applied to atoms in the interior of ligaments to enhance visualization of dislocations. D. Shear strain distribution at a penetration depth of 6 nm. Arrows indicate planar shear localization, consistent with dislocation activity. Bulkier shear distribution (white ellipse) corresponds to a twin, as evidenced by the significant lattice rotation (blue V-shape) within a ligament (green atoms). The gray background structure indicates the undeformed foam configuration. Comparison with the deformed state reveals a minimal increase in solid fraction.
}
\label{fig:MDResults}
\end{figure}

Fig. \ref{fig:MDResults}.A-D summarize the key findings of the simulated nanoindentation. A detailed analysis of the ligaments reveals that plasticity is governed by the nucleation of glissile $\Vec{b} = \frac{a_0}{2} \langle 111 \rangle$ dislocations at ligament surfaces (Fig. \ref{fig:MDResults}.B). Tracking of the nanoporous network displacement (Fig. \ref{fig:MDResults}.C) shows a displacement profile consistent with previous reports on limited densification \cite{briot2018focused,huber2023densification} and is consistent with our SEM observations (Fig. \ref{fig:NI_curves}.C). Interaction of glissile dislocations can lead to sessile  $\Vec{b} = a_0 \langle 100 \rangle$ dislocations.  Tracking dislocation evolution shows that a significant amount of the dislocations annihilate upon reaching a nearby surface. This is also evident after shear localization, as shown in Fig. \ref{fig:MDResults}.D. Some ligament necks display shear localization with no residual dislocations, a signature of dislocation nucleation, propagation and annihilation at nearby surfaces.  Twinning was also found in the vicinity of the tip (inset), in agreement with previous studies \cite{valencia2022nanoindentation,von2024topological}.  
\REV{The MD simulations necessarily involve ligament diameters much smaller than those measured experimentally, approximately 3.1 nm in the simulations compared with approximately 200 nm in the np-Ta particles. Consequently, the simulations are not intended to quantitatively reproduce the experimental indentation modulus, hardness, or load–displacement response. Instead, they are used as an in-silico mechanistic probe \cite{zepeda2021atomistic} to identify the elementary deformation processes active in np-Ta during indentation. The reduced ligament size may favor dislocation escape and annihilation at nearby free surfaces, whereas the larger experimental ligaments may allow greater dislocation retention, storage, and interaction. Nevertheless, this size effect is not expected to change the qualitative conclusion that indentation plasticity in np-Ta is primarily dislocation-mediated, with localized twinning and limited densification.}
Comparison of the deformed sample against its initial state -gray background on inset- shows minimum  increase in solid fraction, consistent with our SEM observations for similar penetration in terms of depth-to-ligament-size ratio (Fig. \ref{fig:MDResults}.D). Furthermore, the plastic zone extends only 3–4 times the penetration depth (Fig. \ref{fig:MDResults}.D), in agreement with experimental observations on np-Au \cite{briot2018focused}. This further indicates that the bonding agent used in our experiments has a negligible influence on the nanoindentation measurements.  In addition, our results show that except for a small twinning fraction, the deformation behavior is identical to that of np-Au during indentation, suggesting that fundamental understanding gained from np-Au studies \cite{richert2021image, hodge2012mechanical, huber2023densification, sun2009situ,sun2007mechanical} could be applicable to other nanoporous metals. 

\begin{figure}[H]
\centering
\includegraphics[width=1.0\textwidth]{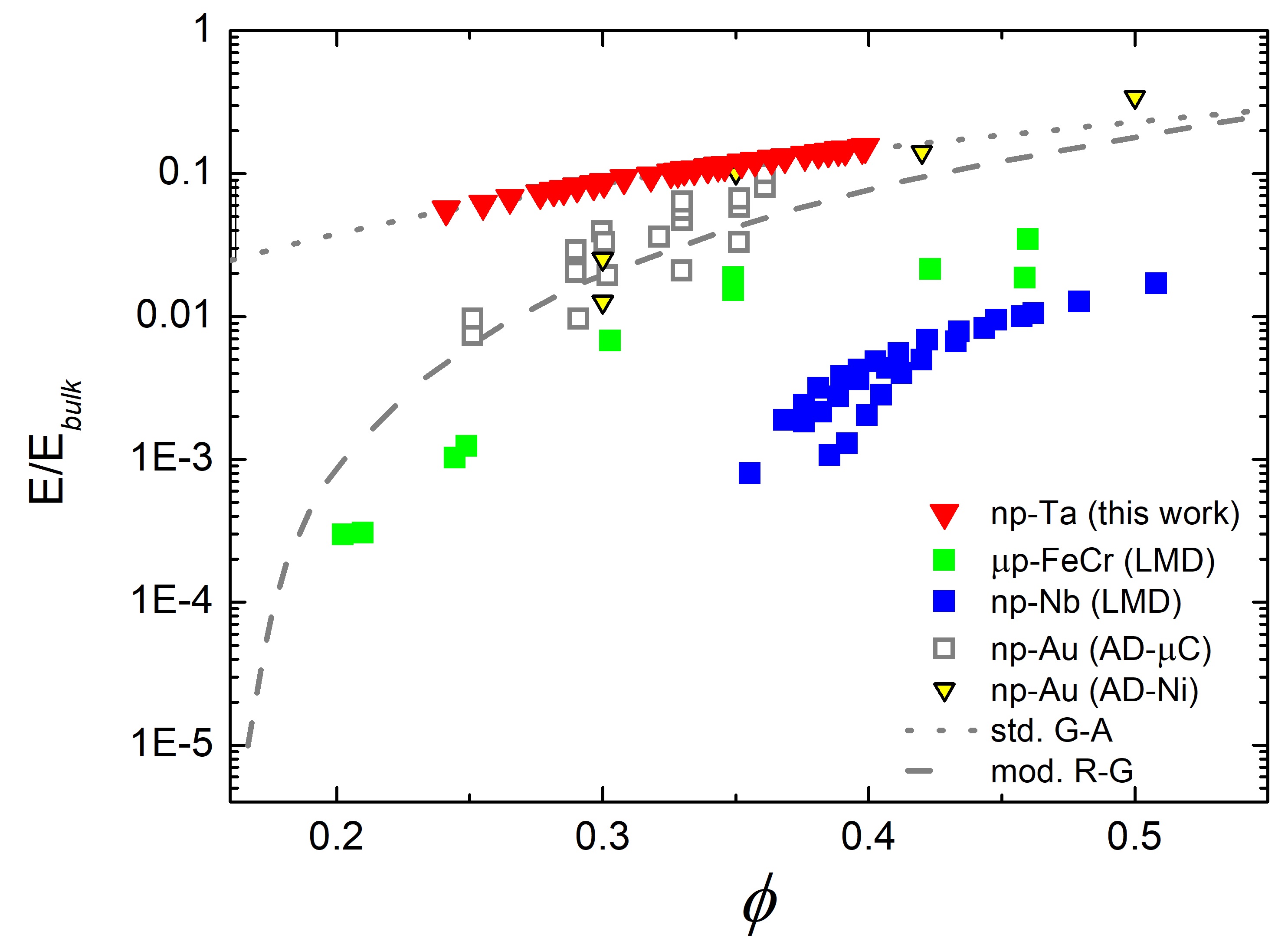}
\caption{Scaled elastic modulus. np-Ta results (this work) are scattered in a region close to the predicted by a standard Gibson-Ashby (std G-A) equation. Triangle symbols indicate nanoindentation results, whereas square symbols indicate micro-compression and/or micro-tension results. For comparison purposes, data corresponding to np-Nb \cite{sohn2025compressive} and $\mu p$-FeCr  \cite{xiang2020universal} produced by LMD and np-Au produced by aqueous dealloying (AD) are also included. np-Au nanoindentation (AD-Ni) data after \cite{biener2005nanoporous,hodge2009ag,huber2023densification} and np-Au micromechanical tests (AD-$\mu$C) after \cite{volkert2006approaching,balk2009tensile,briot2014mechanical,mameka2016nanoporous,burckert2017uniaxial},
together with a modified Roberts-Garboczi (mod. R-G) \cite{soyarslan20183d}. Data points are scaled by the elastic modulus of the corresponding bulk material ($E_{bulk}$).}\label{fig:ScaledE}
\end{figure}

To analyze the measured stiffness in the context of reported scaling trends across nanoporous metals, Fig. \ref{fig:ScaledE} displays the dependence of the elastic modulus of np-Ta normalized by the elastic modulus of bulk Ta (E$_{bulk}$ $\approx$ 186 GPa \cite{ruestes2014atomistic}) compared to experimental measurements on np-metals using nanoindentation and uniaxial microscale tests. While indentation and uniaxial microscale tests involve distinct stress states, Fig.~\ref{fig:ScaledE} is used to contextualize our results against published trends for np-metals. Fig. \ref{fig:ScaledE} further highlights that the np-Ta presented here obeys, to a first approximation, Gibson-Ashby-type scaling laws. These scaling laws were originally developed for porous materials with constant connectivity, which has been poor at describing nanoporous metals because they typically display a solid-fraction dependent connectivity \cite{liu2016interpreting,zandersons2021factors};  a modified Roberts-Garboczi equation (see SM) has had better success in capturing experimental trends; this model is also presented in Fig. \ref{fig:ScaledE} for reference.

Combining the MD observations with the nanoindentation results, it appears that, aside from a small and localized fraction of twinning , no unusual deformation mechanisms are active in the np-Ta ligaments that could explain the trends observed in Figs. \ref{fig:Histo} and \ref{fig:ScaledE}. 
As a result, the comparatively good agreement with Gibson--Ashby-type scaling is plausibly associated with an architecture in which load transfer is not strongly disrupted by disconnected ligaments, i.e., with enhanced load-bearing connectivity. 

Indeed, the SEM images (Fig. \ref{fig:Sample}.B) suggest a bicontinuous, qualitatively well-interconnected ligament morphology, with few apparent isolated termini and numerous visible junctions. \REV{However, because the present study does not include 3D tomography, skeletonization, genus analysis, or effective load-bearing solid-fraction measurements, \cite{chen2012structural,lilleodden2018topological} this connectivity assessment remains qualitative. The mechanical response is therefore interpreted as being consistent with enhanced load-bearing connectivity, rather than as direct quantitative proof of increased topological connectivity.}  This connectivity appears qualitatively higher not only than that reported previously for np-Au, but also than that reported for np-Nb and $\mu$p-FeCr fabricated via LMD. This is particularly surprising because the connectivity in LMD porous structures tends to be less than those fabricated with electrochemical dealloying; this decreased connectivity is attributed to the faster dissolution rate in LMD and transition from interface-limited kinetics to diffusion-limited kinetics for the velocity of the dissolution front \cite{mccue2018pattern}. These differences are clearly apparent in Fig. \ref{fig:ScaledE}, where the literature results for electrochemically dealloyed np-Au  \cite{biener2005nanoporous,volkert2006approaching,balk2009tensile,hodge2009ag,briot2014mechanical,mameka2016nanoporous,burckert2017uniaxial} are broadly distributed following a modified Roberts-Garboczi model trend \cite{soyarslan20183d} whereas the LMD literature results exhibit even poorer scaling. This observation motivates the systematic evaluation of connectivity-aware scaling formulations (e.g., effective load-bearing fraction or genus-based prefactors) for LMD-derived nanoporous metals across different compositions and processing conditions. \REV{Quantitative 3D connectivity metrics and independent density calibration would be valuable in future work to further refine the confidence bounds on $E(\phi)$ and to enable a systematic assessment of connectivity-aware scaling models  \cite{liu2016interpreting,liu2017scaling,sohn2024scaling}.}

The key to the \REV{apparent} high degree of connectivity in np-Ta can be linked to the dealloying method and solvent used in this work. For a fixed parent alloy composition, it was noted in  \cite{lai2022topological} that the topological connectivity of a nanoporous material fabricated via LMD is always lower than one produced electrochemically. The origin of this difference is the diffusion-coupled growth mechanism of the porous structure, which is highly sensitive to “leak” of the remaining/immiscible component (here, Ta) that will be present as a small fraction in the molten metal at the solid/liquid interface during dealloying. However, a benefit of LMD is that the chemistry of the liquid metal solvent can be tuned to alter the solubility of Ta and generate a more connected structure. While the study in \cite{lai2022topological}  examined the effect of Ag additions to molten Cu, it is not surprising that Bi has a similar effect given its binary phase diagram with Ta. The LMD studies included in Fig. \ref{fig:ScaledE} used pure Mg as the molten metal solvent, rather than a tailored melt chemistry, which could explain the scaling differences between \cite{wada2013three,sohn2025compressive} and this work.

\section{Conclusions}

This study reported on the mechanical behavior of single-crystalline nanoporous tantalum produced by liquid metal dealloying of Ti$_{65}$Ta$_{35}$ foils in molten Cu$_{40}$Bi$_{60}$. The specific conclusions are as follows:

\begin{itemize}
    \item Nanoindentation of individual microparticles immobilized on Si substrates reported elastic moduli of 10–30 GPa and hardnesses of 0.3–1.1 GPa at a solid volume fraction of approx. 0.35.
    \item The E values correspond to normalized moduli $E/E_{bulk}$ well above those reported for LMD-derived np-Nb and $\mu$p-FeCr at comparable densities.
    \item The distributions of E and H are consistent with Gibson–Ashby-type scaling once specimen-to-specimen variations in solid fraction are accounted for.
    \item Molecular dynamics simulations of nanoindentation showed that plasticity in np-Ta is dominated by surface nucleation, glide, and annihilation of ⟨111⟩ dislocations, with a small and localized twinning contribution near the contact region.
    \item The nanoscale deformation behavior is qualitatively similar to that reported for np-Au and does not account for the differences in scaling between np-Ta and prior reports on nanoporous metals.
    \item Compared to np-Au and np-Nb, the np-Ta particles exhibit a qualitatively well-interconnected ligament morphology, which is consistent with recent studies that the solubility of Ti and Ta in CuBi molten baths plays a large role on the morphological evolution of the nanoporous network.
\end{itemize}

These findings carry two broader implications. First, mechanical scaling trends developed for electrochemically dealloyed systems such as np-Au are not directly transferable to LMD-derived structures, and benchmarking against the appropriate dealloying route is essential when designing nanoporous metals for a target mechanical response. Second, solvent chemistry in LMD is not solely a knob to control the ligament diameter, but can affect the underlying morphology and ligament connectivity. 
\REV{The present results establish this effect in Ta dealloyed from Ti–Ta in Cu–Bi, but these effects should readily translate to any refractory element that forms a solid solution with Ti and is immiscible with Cu and Bi (e.g., Nb, Mo, W, V); however, it is possible higher order interaction terms could alter phase stability and this trend could vary in system-specific ways.}
\REV{Future work directly quantifying 3D connectivity — via tomography, genus-based measures, and connectivity-aware scaling formulations — across LMD systems produced in different melt chemistries would enable a systematic test of this design strategy and refine the scaling relations that connect synthesis, topology, and mechanical response in refractory nanoporous metals.}

\section*{Acknowledgements}
J.I. Ramallo acknowledges support from CONICET through a postdoctoral scholarship. C.J. Ruestes acknowledges support from the Spanish Ministry of Science and Innovation through a Ramón y Cajal RYC2022-035570-I fellowship and project PID2023-149089OB-I00 (BETMASFUS). I. McCue acknowledges support by the U.S. Department of Energy (DOE), Office of Science, Fusion Energy Sciences, under contract DE-SC0023219. M. C. Fuertes
acknowledges support from CONICET (PIP 11220210100158CO). The authors would like to thank Paula Alderete for performing SEM and EDS measurements, Juan Pablo Coronel for providing the XRD patterns, and Manuel Avella for SEM equipment calibration at IMDEA Materials Institute.

\section*{CRediT authorship contribution statement}
\textbf{Juan Ignacio Ramallo:} Data curation, Formal analysis, Investigation, Methodology, Software, Visualization, Writing - original draft, Writing - review and editing.
\textbf{Nicolás Vázquez von Bibow:} Data curation, Formal analysis, Investigation, Methodology, Software, Visualization.
\textbf{Miguel Monclús:} Formal analysis, Investigation, Resources, Validation.
\textbf{Ian McCue:} Conceptualization, Funding acquisition, Methodology, Resources, Supervision, Writing - review and editing.
\textbf{María Cecilia Fuertes:} Conceptualization, Formal analysis, Methodology, Project administration, Resources, Supervision, Writing - original draft, Writing - review and editing.
\textbf{Carlos Javier Ruestes:} Conceptualization, Funding acquisition, Methodology, Project administration, Supervision, Writing - original draft, Writing - review and editing.

\section*{Conflict of Interest}
The authors declare no conflicts of interest.

\bibliographystyle{elsarticle-num} 
\bibliography{cas-refs}

@article{chuang2022powder,
  title={A powder metallurgy approach to liquid metal dealloying with applications in additive manufacturing},
  author={Chuang, A and Baris, J and Ott, C and McCue, I and Erlebacher, J},
  journal={Acta Materialia},
  volume={238},
  pages={118213},
  year={2022},
  publisher={Elsevier}
}

@article{stuckner2017aquami,
  title={AQUAMI: An open source Python package and GUI for the automatic quantitative analysis of morphologically complex multiphase materials},
  author={Stuckner, Joshua and Frei, Katherine and McCue, Ian and Demkowicz, Michael J and Murayama, Mitsuhiro},
  journal={Computational Materials Science},
  volume={139},
  pages={320--329},
  year={2017},
  publisher={Elsevier}
}

@article{ruestes2014atomistic,
  title={Atomistic simulation of tantalum nanoindentation: Effects of indenter diameter, penetration velocity, and interatomic potentials on defect mechanisms and evolution},
  author={Ruestes, Carlos Javier and Stukowski, Alexander and Tang, Yizhe and Tramontina, DR and Erhart, Paul and Remington, BA and Urbassek, HM and Meyers, Marc A and Bringa, Eduardo Marcial},
  journal={Materials Science and Engineering: A},
  volume={613},
  pages={390--403},
  year={2014},
  publisher={Elsevier}
}

@article{sohn2025compressive,
  title={Compressive behavior and connecting topology of monolithic nanoporous niobium},
  author={Sohn, Seoyun and Shi, Shan and Markmann, J{\"u}rgen and Berger, Stefan Alexander and Weissm{\"u}ller, J{\"o}rg},
  journal={Materials Research Letters},
  volume={13},
  number={1},
  pages={76--85},
  year={2025},
  publisher={Taylor \& Francis}
}

@article{briot2014mechanical,
  title={Mechanical properties of bulk single crystalline nanoporous gold investigated by millimetre-scale tension and compression testing},
  author={Briot, Nicolas J and Kennerknecht, Tobias and Eberl, Christoph and Balk, T John},
  journal={Philosophical Magazine},
  volume={94},
  number={8},
  pages={847--866},
  year={2014},
  publisher={Taylor \& Francis}
}

@article{volkert2006approaching,
  title={Approaching the theoretical strength in nanoporous Au},
  author={Volkert, CA and Lilleodden, ET and Kramer, D and Weissm{\"u}ller, J},
  journal={Applied Physics Letters},
  volume={89},
  number={6},
  year={2006},
  publisher={AIP Publishing}
}

@article{biener2005nanoporous,
  title={Nanoporous Au: A high yield strength material},
  author={Biener, Juergen and Hodge, Andrea M and Hamza, Alex V and Hsiung, Luke M and Satcher, Joe H},
  journal={Journal of applied physics},
  volume={97},
  number={2},
  year={2005},
  publisher={AIP Publishing}
}

@article{balk2009tensile,
  title={Tensile and compressive microspecimen testing of bulk nanoporous gold},
  author={Balk, T John and Eberl, Chris and Sun, Ye and Hemker, Kevin J and Gianola, Daniel S},
  journal={Jom},
  volume={61},
  number={12},
  pages={26--31},
  year={2009},
  publisher={Springer}
}

@article{hodge2009ag,
  title={Ag effects on the elastic modulus values of nanoporous Au foams},
  author={Hodge, AM and Doucette, RT and Biener, MM and Biener, J and Cervantes, O and Hamza, AV},
  journal={Journal of Materials Research},
  volume={24},
  number={4},
  pages={1600--1606},
  year={2009},
  publisher={Cambridge University Press}
}

@article{burckert2017uniaxial,
  title={Uniaxial compression testing of bulk nanoporous gold},
  author={B{\"u}rckert, Michael and Briot, Nicolas J and Balk, T John},
  journal={Philosophical Magazine},
  volume={97},
  number={15},
  pages={1157--1178},
  year={2017},
  publisher={Taylor \& Francis}
}

@article{mameka2016nanoporous,
  title={Nanoporous gold—testing macro-scale samples to probe small-scale mechanical behavior},
  author={Mameka, Nadiia and Wang, Ke and Markmann, J{\"u}rgen and Lilleodden, Erica T and Weissm{\"u}ller, J{\"o}rg},
  journal={Materials Research Letters},
  volume={4},
  number={1},
  pages={27--36},
  year={2016},
  publisher={Taylor \& Francis}
}

@article{gibson1982mechanics,
  title={The mechanics of three-dimensional cellular materials},
  author={Gibson, IJ and Ashby, Michael Farries},
  journal={Proceedings of the royal society of London. A. Mathematical and physical sciences},
  volume={382},
  number={1782},
  pages={43--59},
  year={1982},
  publisher={The Royal Society London}
}

@article{soyarslan20183d,
  title={3D stochastic bicontinuous microstructures: Generation, topology and elasticity},
  author={Soyarslan, Celal and Bargmann, Swantje and Pradas, Marc and Weissm{\"u}ller, J{\"o}rg},
  journal={Acta materialia},
  volume={149},
  pages={326--340},
  year={2018},
  publisher={Elsevier}
}

@article{liu2016interpreting,
  title={Interpreting anomalous low-strength and low-stiffness of nanoporous gold: Quantification of network connectivity},
  author={Liu, Ling-Zhi and Ye, Xing-Long and Jin, Hai-Jun},
  journal={Acta Materialia},
  volume={118},
  pages={77--87},
  year={2016},
  publisher={Elsevier}
}

@article{zepeda2021atomistic,
  title={Atomistic insights into metal hardening},
  author={Zepeda-Ruiz, Luis A and Stukowski, Alexander and Oppelstrup, Tomas and Bertin, Nicolas and Barton, Nathan R and Freitas, Rodrigo and Bulatov, Vasily V},
  journal={Nature materials},
  volume={20},
  number={3},
  pages={315--320},
  year={2021},
  publisher={Nature Publishing Group UK London}
}

@article{briot2018focused,
  title={Focused ion beam characterization of deformation resulting from nanoindentation of nanoporous gold},
  author={Briot, Nicolas J and Balk, T John},
  journal={MRS Communications},
  volume={8},
  number={1},
  pages={132--136},
  year={2018},
  publisher={Cambridge University Press}
}

@article{von2024topological,
  title={Topological changes and deformation mechanisms of nanoporous Ta under compression},
  author={Von Bibow, N Vazquez and Mill{\'a}n, Emmanuel Nicol{\'a}s and Ruestes, Carlos Javier},
  journal={Computational Materials Science},
  volume={236},
  pages={112884},
  year={2024},
  publisher={Elsevier}
}

@article{valencia2022nanoindentation,
  title={Nanoindentation of nanoporous tungsten: A molecular dynamics approach},
  author={Valencia, Felipe J and Ortega, Robinson and Gonz{\'a}lez, Rafael I and Bringa, Eduardo M and Kiwi, Miguel and Ruestes, Carlos J},
  journal={Computational Materials Science},
  volume={209},
  pages={111336},
  year={2022},
  publisher={Elsevier}
}

@article{biener2009surface,
  title={Surface-chemistry-driven actuation in nanoporous gold},
  author={Biener, J and Wittstock, A and Zepeda-Ruiz, LA and Biener, MM and Zielasek, V and Kramer, D and Viswanath, RN and Weissm{\"u}ller, J and B{\"a}umer, M and Hamza, AV},
  journal={Nature materials},
  volume={8},
  number={1},
  pages={47--51},
  year={2009},
  publisher={Nature Publishing Group UK London}
}

@article{detsi2016metallic,
  title={Metallic muscles and beyond: nanofoams at work},
  author={Detsi, Eric and Tolbert, Sarah H and Punzhin, S and De Hosson, Jeff Th M},
  journal={Journal of materials science},
  volume={51},
  number={1},
  pages={615--634},
  year={2016},
  publisher={Springer}
}

@article{zonana2020effect,
  title={Effect of tip roundness on the nanoindentation of Fe crystals},
  author={Zonana, M Clara and Ruestes, Carlos J and Bringa, Eduardo M and Urbassek, Herbert M},
  journal={Tribology Letters},
  volume={68},
  number={2},
  pages={56},
  year={2020},
  publisher={Springer}
}

@article{janisch2025nanoindentation,
  title={The nanoindentation puzzle: Putting the pieces together using simulations at different scales: R. Janisch, D. Mordehai},
  author={Janisch, Rebecca and Mordehai, Dan},
  journal={MRS Bulletin},
  pages={1--14},
  year={2025},
  publisher={Springer}
}

@article{ruestes2017molecular,
  title={Molecular dynamics modeling of nanoindentation},
  author={Ruestes, Carlos J and Bringa, Eduardo M and Gao, Yu and Urbassek, Herbert M},
  journal={Applied nanoindentation in advanced materials},
  pages={313--345},
  year={2017},
  publisher={Wiley Online Library}
}

@article{fischer2004nanoindentation,
  title={Nanoindentation. Mechanical engineering series},
  author={Fischer-Cripps, Anthony C and Nicholson, DW},
  journal={Appl. Mech. Rev.},
  volume={57},
  number={2},
  pages={B12--B12},
  year={2004}
}

@article{alcala2012planar,
  title={Planar Defect Nucleation and Annihilation Mechanisms in Nanocontact Plasticity of Metal Surfaces},
  author={Alcal{\'a}, Jorge and Dalmau, Roger and Franke, Oliver and Biener, Monika and Biener, Juergen and Hodge, Andrea},
  journal={Physical Review Letters},
  volume={109},
  number={7},
  pages={075502},
  year={2012},
  publisher={APS}
}

@article{thompson2022lammps,
  title={LAMMPS-a flexible simulation tool for particle-based materials modeling at the atomic, meso, and continuum scales},
  author={Thompson, Aidan P and Aktulga, H Metin and Berger, Richard and Bolintineanu, Dan S and Brown, W Michael and Crozier, Paul S and In't Veld, Pieter J and Kohlmeyer, Axel and Moore, Stan G and Nguyen, Trung Dac and others},
  journal={Computer physics communications},
  volume={271},
  pages={108171},
  year={2022},
  publisher={Elsevier}
}

@article{stukowski2009visualization,
  title={Visualization and analysis of atomistic simulation data with OVITO--the Open Visualization Tool},
  author={Stukowski, Alexander},
  journal={Modelling and simulation in materials science and engineering},
  volume={18},
  number={1},
  pages={015012},
  year={2009},
  publisher={IOP Publishing}
}

@article{stukowski2010extracting,
  title={Extracting dislocations and non-dislocation crystal defects from atomistic simulation data},
  author={Stukowski, Alexander and Albe, Karsten},
  journal={Modelling and Simulation in Materials Science and Engineering},
  volume={18},
  number={8},
  pages={085001},
  year={2010},
  publisher={IOP Publishing}
}

@article{ruestes2025nanoporous,
  title={Nanoporous metals under extremes},
  author={Ruestes, Carlos J and Farkas, Diana and Snyder, Joshua},
  journal={MRS bulletin},
  pages={1--11},
  year={2025},
  publisher={Springer}
}

@article{shi2021scaling,
  title={Scaling behavior of stiffness and strength of hierarchical network nanomaterials},
  author={Shi, Shan and Li, Yong and Ngo-Dinh, Bao-Nam and Markmann, J{\"u}rgen and Weissm{\"u}ller, J{\"o}rg},
  journal={Science},
  volume={371},
  number={6533},
  pages={1026--1033},
  year={2021},
  publisher={American Association for the Advancement of Science}
}

@article{zandersons2021factors,
  title={On factors defining the mechanical behavior of nanoporous gold},
  author={Zandersons, Birthe and L{\"u}hrs, Lukas and Li, Yong and Weissm{\"u}ller, J{\"o}rg},
  journal={Acta materialia},
  volume={215},
  pages={116979},
  year={2021},
  publisher={Elsevier}
}

@article{engelman2025deformation,
  title={Deformation behavior of nanoporous gold nanoparticles during compression},
  author={Engelman, Ben and Mathesan, Santhosh and Fedyaeva, Tatyana and Bisht, Anuj and Rabkin, Eugen and Mordehai, Dan},
  journal={Acta Materialia},
  volume={286},
  pages={120723},
  year={2025},
  publisher={Elsevier}
}

@article{liu2017scaling,
  title={Scaling equation for the elastic modulus of nanoporous gold with “fixed” network connectivity},
  author={Liu, Ling-Zhi and Jin, Hai-Jun},
  journal={Applied Physics Letters},
  volume={110},
  number={21},
  year={2017},
  publisher={AIP Publishing}
}

@article{wittstock2023nanoporous,
  title={Nanoporous gold: from structure evolution to functional properties in catalysis and electrochemistry},
  author={Wittstock, Gunther and B{\"a}umer, Marcus and Dononelli, Wilke and Kl{\"u}ner, Thorsten and L{\"u}hrs, Lukas and Mahr, Christoph and Moskaleva, Lyudmila V and Oezaslan, Mehtap and Risse, Thomas and Rosenauer, Andreas and others},
  journal={Chemical reviews},
  volume={123},
  number={10},
  pages={6716--6792},
  year={2023},
  publisher={ACS Publications}
}

@article{zhang2013nanoporous,
  title={Nanoporous gold based optical sensor for sub-ppt detection of mercury ions},
  author={Zhang, Ling and Chang, Haixin and Hirata, Akihiko and Wu, Hongkai and Xue, Qi-Kun and Chen, Mingwei},
  journal={ACS nano},
  volume={7},
  number={5},
  pages={4595--4600},
  year={2013},
  publisher={ACS Publications}
}

@article{bringa2012nanoporous,
  title={Are nanoporous materials radiation resistant?},
  author={Bringa, Eduardo Marcial and Monk, JD and Caro, A and Misra, A and Zepeda-Ruiz, L and Duchaineau, M and Abraham, F and Nastasi, Michael and Picraux, Samuel Thomas and Wang, YQ and others},
  journal={Nano letters},
  volume={12},
  number={7},
  pages={3351--3355},
  year={2012},
  publisher={ACS Publications}
}

@article{mccue2018pattern,
  title={Pattern formation during electrochemical and liquid metal dealloying},
  author={McCue, Ian and Karma, Alain and Erlebacher, Jonah},
  journal={Mrs Bulletin},
  volume={43},
  number={1},
  pages={27--34},
  year={2018},
  publisher={Cambridge University Press}
}

@article{wada2011dealloying,
  title={Dealloying by metallic melt},
  author={Wada, Takeshi and Yubuta, Kunio and Inoue, Akihisa and Kato, Hidemi},
  journal={Materials Letters},
  volume={65},
  number={7},
  pages={1076--1078},
  year={2011},
  publisher={Elsevier}
}

@article{wada2013three,
  title={Three-dimensional open-cell macroporous iron, chromium and ferritic stainless steel},
  author={Wada, Takeshi and Kato, Hidemi},
  journal={Scripta Materialia},
  volume={68},
  number={9},
  pages={723--726},
  year={2013},
  publisher={Elsevier}
}

@article{geslin2015topology,
  title={Topology-generating interfacial pattern formation during liquid metal dealloying},
  author={Geslin, Pierre-Antoine and McCue, Ian and Gaskey, Bernard and Erlebacher, Jonah and Karma, Alain},
  journal={Nature communications},
  volume={6},
  number={1},
  pages={8887},
  year={2015},
  publisher={Nature Publishing Group UK London}
}

@article{mccue2016kinetics,
  title={Kinetics and morphological evolution of liquid metal dealloying},
  author={McCue, Ian and Gaskey, Bernard and Geslin, Pierre-Antoine and Karma, Alain and Erlebacher, Jonah},
  journal={Acta Materialia},
  volume={115},
  pages={10--23},
  year={2016},
  publisher={Elsevier}
}

@article{kim2014sub,
  title={Sub-micron porous niobium solid electrolytic capacitor prepared by dealloying in a metallic melt},
  author={Kim, Joung Wook and Wada, Takeshi and Kim, Sung Gyoo and Kato, Hidemi},
  journal={Materials Letters},
  volume={116},
  pages={223--226},
  year={2014},
  publisher={Elsevier}
}

@article{gaskey2019self,
  title={Self-assembled porous metal-intermetallic nanocomposites via liquid metal dealloying},
  author={Gaskey, Bernard and McCue, Ian and Chuang, Alyssa and Erlebacher, Jonah},
  journal={Acta Materialia},
  volume={164},
  pages={293--300},
  year={2019},
  publisher={Elsevier}
}

@article{okulov2017dealloying,
  title={Dealloying-based interpenetrating-phase nanocomposites matching the elastic behavior of human bone},
  author={Okulov, IV and Weissm{\"u}ller, J{\"o}rg and Markmann, J{\"u}rgen},
  journal={Scientific reports},
  volume={7},
  number={1},
  pages={20},
  year={2017},
  publisher={Nature Publishing Group UK London}
}

@article{berger2020open,
  title={Open porous $\alpha$+ $\beta$ titanium alloy by liquid metal dealloying for biomedical applications},
  author={Berger, Stefan Alexander and Okulov, Ilya Vladimirovich},
  journal={Metals},
  volume={10},
  number={11},
  pages={1450},
  year={2020},
  publisher={MDPI}
}

@article{xiang2020universal,
  title={A universal scaling relationship between the strength and Young’s modulus of dealloyed porous Fe0. 80Cr0. 20},
  author={Xiang, Yi-Hou and Liu, Ling-Zhi and Shao, Jun-Chao and Jin, Hai-Jun},
  journal={Acta Materialia},
  volume={186},
  pages={105--115},
  year={2020},
  publisher={Elsevier}
}

@article{zhao2017three,
  title={Three-dimensional morphological and chemical evolution of nanoporous stainless steel by liquid metal dealloying},
  author={Zhao, Chonghang and Wada, Takeshi and De Andrade, Vincent and Williams, Garth J and Gelb, Jeff and Li, Li and Thieme, Juergen and Kato, Hidemi and Chen-Wiegart, Yu-chen Karen},
  journal={ACS applied materials \& interfaces},
  volume={9},
  number={39},
  pages={34172--34184},
  year={2017},
  publisher={ACS Publications}
}

@article{okulov2020nanoporous,
  title={Nanoporous high-entropy alloy by liquid metal dealloying},
  author={Okulov, Artem Vladimirovich and Joo, Soo-Hyun and Kim, Hyoung Seop and Kato, Hidemi and Okulov, Ilya Vladimirovich},
  journal={Metals},
  volume={10},
  number={10},
  pages={1396},
  year={2020},
  publisher={MDPI}
}

@article{lai2022topological,
  title={Topological control of liquid-metal-dealloyed structures},
  author={Lai, Longhai and Gaskey, Bernard and Chuang, Alyssa and Erlebacher, Jonah and Karma, Alain},
  journal={Nature Communications},
  volume={13},
  number={1},
  pages={2918},
  year={2022},
  publisher={Nature Publishing Group UK London}
}

@article{wada2011nano,
  title={Nano-to submicro-porous $\beta$-Ti alloy prepared from dealloying in a metallic melt},
  author={Wada, Takeshi and Setyawan, Albertus Deny and Yubuta, Kunio and Kato, Hidemi},
  journal={Scripta Materialia},
  volume={65},
  number={6},
  pages={532--535},
  year={2011},
  publisher={Elsevier}
}

@article{wada2024accelerated,
  title={Accelerated structural ordering during liquid metal dealloying and its effect on thermal coarsening of nanoporous refractory alloys},
  author={Wada, Takeshi and Nakata, Akira and Song, Ruirui and Kato, Hidemi},
  journal={Scripta Materialia},
  volume={247},
  pages={116120},
  year={2024},
  publisher={Elsevier}
}

@article{chakraborti2016ultrathin,
  title={Ultrathin, substrate-integrated, and self-healing nanocapacitors with low-leakage currents and high-operating frequencies},
  author={Chakraborti, Parthasarathi and Sharma, Himani and Pulugurtha, Markondeya Raj and Rataj, Kamil-Paul and Schnitter, Christopher and Neuhart, Nathan and Jain, Shubham and Gandhi, Saumya and Tummala, Rao R},
  journal={IEEE Transactions on Components, Packaging and Manufacturing Technology},
  volume={6},
  number={12},
  pages={1776--1784},
  year={2016},
  publisher={IEEE}
}

@article{wauthle2015additively,
  title={Additively manufactured porous tantalum implants},
  author={Wauthle, Ruben and Van der Stok, Johan and Yavari, Saber Amin and Van Humbeeck, Jan and Kruth, Jean-Pierre and Zadpoor, Amir Abbas and Weinans, Harrie and Mulier, Michiel and Schrooten, Jan},
  journal={Acta biomaterialia},
  volume={14},
  pages={217--225},
  year={2015},
  publisher={Elsevier}
}

@article{oliver1992improved,
  title={An improved technique for determining hardness and elastic modulus using load and displacement sensing indentation experiments},
  author={Oliver, Warren Carl and Pharr, George Mathews},
  journal={Journal of materials research},
  volume={7},
  number={6},
  pages={1564--1583},
  year={1992},
  publisher={Cambridge University Press}
}

@article{luhrs2016elastic,
  title={Elastic and plastic Poisson’s ratios of nanoporous gold},
  author={L{\"u}hrs, Lukas and Soyarslan, Celal and Markmann, J{\"u}rgen and Bargmann, Swantje and Weissm{\"u}ller, J{\"o}rg},
  journal={Scripta Materialia},
  volume={110},
  pages={65--69},
  year={2016},
  publisher={Elsevier}
}

@article{lilleodden2018topological,
  title={On the topological, morphological, and microstructural characterization of nanoporous metals},
  author={Lilleodden, Erica T and Voorhees, Peter W},
  journal={MRS bulletin},
  volume={43},
  number={1},
  pages={20--26},
  year={2018},
  publisher={Cambridge University Press}
}

@article{sohn2024scaling,
  title={Scaling between elasticity and topological genus for random network nanomaterials},
  author={Sohn, Seoyun and Richert, Claudia and Shi, Shan and Weissm{\"u}ller, J{\"o}rg and Huber, Norbert},
  journal={Extreme mechanics letters},
  volume={68},
  pages={102147},
  year={2024},
  publisher={Elsevier}
}

@article{ruestes2016hardening,
  title={Hardening under compression in Au foams},
  author={Ruestes, Carlos J and Farkas, Diana and Caro, Alfredo and Bringa, Eduardo M},
  journal={Acta Materialia},
  volume={108},
  pages={1--7},
  year={2016},
  publisher={Elsevier}
}

@article{ngo2015anomalous,
  title={Anomalous compliance and early yielding of nanoporous gold},
  author={Ng{\^o}, Bao-Nam Dinh and Stukowski, Alexander and Mameka, Nadiia and Markmann, J{\"u}rgen and Albe, Karsten and Weissm{\"u}ller, J{\"o}rg},
  journal={Acta Materialia},
  volume={93},
  pages={144--155},
  year={2015},
  publisher={Elsevier}
}

@article{montemayor2015materials,
  title={Materials by design: Using architecture in material design to reach new property spaces},
  author={Montemayor, Lauren and Chernow, Victoria and Greer, Julia R},
  journal={Mrs Bulletin},
  volume={40},
  number={12},
  pages={1122--1129},
  year={2015},
  publisher={Cambridge University Press}
}

@article{greer2019three,
  title={Three-dimensional architected materials and structures: Design, fabrication, and mechanical behavior},
  author={Greer, Julia R and Deshpande, Vikram S},
  journal={MRS Bulletin},
  volume={44},
  number={10},
  pages={750--757},
  year={2019},
  publisher={Cambridge University Press}
}

@article{surjadi2025double,
  title={Double-network-inspired mechanical metamaterials},
  author={Surjadi, James Utama and Aymon, Bastien FG and Carton, Molly and Portela, Carlos M},
  journal={Nature Materials},
  pages={1--10},
  year={2025},
  publisher={Nature Publishing Group UK London}
}

@article{surjadi2025enabling,
  title={Enabling three-dimensional architected materials across length scales and timescales},
  author={Surjadi, James Utama and Portela, Carlos M},
  journal={Nature Materials},
  pages={1--13},
  year={2025},
  publisher={Nature Publishing Group UK London}
}

@article{vyatskikh2018additive,
  title={Additive manufacturing of 3D nano-architected metals},
  author={Vyatskikh, Andrey and Delalande, St{\'e}phane and Kudo, Akira and Zhang, Xuan and Portela, Carlos M and Greer, Julia R},
  journal={Nature communications},
  volume={9},
  number={1},
  pages={593},
  year={2018},
  publisher={Nature Publishing Group UK London}
}

@article{zhang2021recent,
  title={Recent trends on studies of nanostructured metals},
  author={Zhang, Xinghang and Lilleodden, Erica and Wang, Jian},
  journal={MRS Bulletin},
  volume={46},
  number={3},
  pages={217--224},
  year={2021},
  publisher={Springer}
}

@article{portela2020extreme,
  title={Extreme mechanical resilience of self-assembled nanolabyrinthine materials},
  author={Portela, Carlos M and Vidyasagar, A and Kr{\"o}del, Sebastian and Weissenbach, Tamara and Yee, Daryl W and Greer, Julia R and Kochmann, Dennis M},
  journal={Proceedings of the National Academy of Sciences},
  volume={117},
  number={11},
  pages={5686--5693},
  year={2020},
  publisher={National Academy of Sciences}
}

@article{shaw2019computationally,
  title={Computationally efficient design of directionally compliant metamaterials},
  author={Shaw, Lucas A and Sun, Frederick and Portela, Carlos M and Barranco, Rodolfo I and Greer, Julia R and Hopkins, Jonathan B},
  journal={Nature communications},
  volume={10},
  number={1},
  pages={291},
  year={2019},
  publisher={Nature Publishing Group UK London}
}

@article{wu2023consequences,
  title={On the consequences of intrinsic and extrinsic size effects on the mechanical response of nanoporous Au},
  author={Wu, Yijuan and Markmann, J{\"u}rgen and Lilleodden, Erica T},
  journal={Materials \& Design},
  volume={232},
  pages={112175},
  year={2023},
  publisher={Elsevier}
}

@article{bieberdorf2023grain,
  title={Grain boundary effects in high-temperature liquid-metal dealloying: a multi-phase field study},
  author={Bieberdorf, Nathan and Asta, Mark and Capolungo, Laurent},
  journal={npj computational materials},
  volume={9},
  number={1},
  pages={127},
  year={2023},
  publisher={Nature Publishing Group UK London}
}

@article{kerr2024morphology,
  title={Morphology selection in dealloying: A phase field study of the coupling among kinetic mechanisms},
  author={Kerr, Justin and Bieberdorf, Nathan and Capolungo, Laurent and Asta, Mark},
  journal={Physical Review Materials},
  volume={8},
  number={10},
  pages={103802},
  year={2024},
  publisher={APS}
}

@article{mccue2016dealloying,
  title={Dealloying and dealloyed materials},
  author={McCue, Ian and Benn, Ellen and Gaskey, Bernard and Erlebacher, Jonah},
  journal={Annual review of materials research},
  volume={46},
  number={1},
  pages={263--286},
  year={2016},
  publisher={Annual Reviews}
}

@article{farkas2018indentation,
  title={Indentation response of nanoporous gold from atomistic simulations},
  author={Farkas, Diana and Stuckner, Joshua and Umbel, Rachel and Kuhr, Bryan and Demkowicz, Michael J},
  journal={Journal of Materials Research},
  volume={33},
  number={10},
  pages={1382--1390},
  year={2018},
  publisher={Cambridge University Press}
}

@article{xia2015role,
  title={The role of computer simulation in nanoporous metals—A review},
  author={Xia, Re and Wu, Run Ni and Liu, Yi Lun and Sun, Xiao Yu},
  journal={Materials},
  volume={8},
  number={8},
  pages={5060--5083},
  year={2015},
  publisher={MDPI}
}

@book{ashby2000metal,
  title={Metal foams: a design guide},
  author={Ashby, Michael F},
  year={2000},
  publisher={Elsevier}
}

@article{alhafez2021indentation,
  title={Indentation and scratching of iron by a rotating tool--a molecular dynamics study},
  author={Alhafez, Iyad Alabd and Ruestes, Carlos J and Bringa, Eduardo M and Urbassek, Herbert M},
  journal={Computational Materials Science},
  volume={194},
  pages={110445},
  year={2021},
  publisher={Elsevier}
}

@article{przybilla2025revealing,
  title={Revealing nanoscale plasticity of metallic nanosponges with correlative and scale-bridging 3D microscopy and modelling},
  author={Przybilla, Thomas and Xie, Zhuocheng and Prakash, Aruna and Thiess, Erich and Niekiel, Florian and Apeleo Zubiri, Benjamin and Ma{\v{c}}kovi{\'c}, Mirza and Schweizer, Peter and Gu{\'e}nol{\'e}, Julien and Kelly, Stephen T and others},
  journal={Communications Materials},
  volume={6},
  number={1},
  pages={204},
  year={2025},
  publisher={Nature Publishing Group UK London}
}

@article{garrison2023review,
  title={Review of Recent Progress in Plasma-Facing Material Joints and Composites in the FRONTIER US-Japan Collaboration},
  author={Garrison, Laura M and Katoh, Yuki and Hinoki, Tatsuya and Hashimoto, Naoyuki and Echols, John R and Geringer, Josina W and Reid, Nathan C and Allain, Jean Paul and Cheng, Bin and Dorow-Gerspach, Daniel and others},
  journal={Fusion Science and Technology},
  volume={79},
  number={6},
  pages={662--670},
  year={2023},
  publisher={Taylor \& Francis}
}

@article{peters2024materials,
  title={Materials design for hypersonics},
  author={Peters, Adam B and Zhang, Dajie and Chen, Samuel and Ott, Catherine and Oses, Corey and Curtarolo, Stefano and McCue, Ian and Pollock, Tresa M and Eswarappa Prameela, Suhas},
  journal={Nature communications},
  volume={15},
  number={1},
  pages={3328},
  year={2024},
  publisher={Nature Publishing Group UK London}
}

@article{cunningham2024alloying,
  title={Alloying effects on the microstructure and properties of laser additively manufactured tungsten materials},
  author={Cunningham, W Streit and Lang, Eric and Sprouster, David and Olynik, Nicholas and Pattammattel, Ajith and Olds, Daniel and Hattar, Khalid and McCue, Ian and Trelewicz, Jason R},
  journal={Materials Science and Engineering: A},
  volume={914},
  pages={147110},
  year={2024},
  publisher={Elsevier}
}

@article{mccue2016local,
  title={Local heterogeneity in the mechanical properties of bicontinuous composites made by liquid metal dealloying},
  author={McCue, Ian and Gaskey, Bernard and Crawford, Bryan and Erlebacher, Jonah},
  journal={Applied Physics Letters},
  volume={109},
  number={23},
  year={2016},
  publisher={AIP Publishing}
}

@article{richert2021image,
  title={Image segmentation and analysis for densification mapping of nanoporous gold after nanoindentation},
  author={Richert, Claudia and Wu, Yijuan and Hablitzel, Murilo and Lilleodden, Erica T and Huber, Norbert},
  journal={MRS advances},
  volume={6},
  number={20},
  pages={519--523},
  year={2021},
  publisher={Springer}
}

@article{hodge2012mechanical,
  title={Mechanical properties of nanoporous gold},
  author={Hodge, Andrea M and Balk, Thomas John},
  journal={Nanoporous Gold: From an Ancient Technology to a High-Tech Material, RSC Nanoscience \& Nanotechnology},
  number={22},
  pages={51--68},
  year={2012}
}

@article{huber2023densification,
  title={Densification of nanoporous metals during nanoindentation: The role of structural and mechanical properties},
  author={Huber, Norbert and Ryl, Ilona and Wu, Y and Hablitzel, M and Zandersons, Birthe and Richert, Claudia and Lilleodden, Erica},
  journal={Journal of materials research},
  volume={38},
  number={3},
  pages={853--866},
  year={2023},
  publisher={Springer}
}

@article{sun2007mechanical,
  title={The mechanical behavior of nanoporous gold thin films},
  author={Sun, Ye and Ye, Jia and Shan, Zhiwei and Minor, Andrew M and Balk, T John},
  journal={Jom},
  volume={59},
  number={9},
  pages={54--58},
  year={2007},
  publisher={Springer}
}

@article{sun2009situ,
  title={In situ indentation of nanoporous gold thin films in the transmission electron microscope},
  author={Sun, YE and Ye, JIA and Minor, Andrew M and Balk, T John},
  journal={Microscopy Research and Technique},
  volume={72},
  number={3},
  pages={232--241},
  year={2009},
  publisher={Wiley Online Library}
}

@article{hodge2007scaling,
  title={Scaling equation for yield strength of nanoporous open-cell foams},
  author={Hodge, AM and Biener, J and Hayes, JR and Bythrow, PM and Volkert, CA and Hamza, AV},
  journal={Acta Materialia},
  volume={55},
  number={4},
  pages={1343--1349},
  year={2007},
  publisher={Elsevier}
}

@article{biener2011ald,
  title={ALD functionalized nanoporous gold: thermal stability, mechanical properties, and catalytic activity},
  author={Biener, Monika M and Biener, Juergen and Wichmann, Andre and Wittstock, Arne and Baumann, Theodore F and B{\"a}umer, Marcus and Hamza, Alex V},
  journal={Nano letters},
  volume={11},
  number={8},
  pages={3085--3090},
  year={2011},
  publisher={ACS Publications}
}

@article{kreuzeder2015fabrication,
  title={Fabrication and thermo-mechanical behavior of ultra-fine porous copper},
  author={Kreuzeder, Marius and Abad, Manuel-David and Primorac, Mladen-Mateo and Hosemann, Peter and Maier, Verena and Kiener, Daniel},
  journal={Journal of materials science},
  volume={50},
  number={2},
  pages={634--643},
  year={2015},
  publisher={Springer}
}

@article{leitner2016interface,
  title={Interface dominated mechanical properties of ultra-fine grained and nanoporous Au at elevated temperatures},
  author={Leitner, Alexander and Maier-Kiener, Verena and Jeong, Jiwon and Abad, Manuel D and Hosemann, Peter and Oh, Sang Ho and Kiener, Daniel},
  journal={Acta materialia},
  volume={121},
  pages={104--116},
  year={2016},
  publisher={Elsevier}
}

@article{esque2016nanomechanical,
  title={Nanomechanical behaviour of open-cell nanoporous metals: Homogeneous versus thickness-dependent porosity},
  author={Esqu{\'e}-de los Ojos, Daniel and Zhang, Jin and Fornell, J and Pellicer, E and Sort, J},
  journal={Mechanics of Materials},
  volume={100},
  pages={167--174},
  year={2016},
  publisher={Elsevier}
}

@article{kang2018microstructural,
  title={Microstructural effect on time-dependent plasticity of nanoporous gold},
  author={Kang, Na-Ri and Gwak, Eun-Ji and Jeon, Hansol and Song, Eunji and Kim, Ju-Young},
  journal={International Journal of Plasticity},
  volume={109},
  pages={108--120},
  year={2018},
  publisher={Elsevier}
}

@article{kim2018indentation,
  title={Indentation size effect for spherical nanoindentation on nanoporous gold},
  author={Kim, Young-Cheon and Gwak, Eun-Ji and Ahn, Seung-min and Kang, Na-Ri and Han, Heung Nam and Jang, Jae-il and Kim, Ju-Young},
  journal={Scripta Materialia},
  volume={143},
  pages={10--14},
  year={2018},
  publisher={Elsevier}
}

@article{joo2020beating,
  title={Beating thermal coarsening in nanoporous materials via high-entropy design},
  author={Joo, Soo-Hyun and Bae, Jae Wung and Park, Won-Young and Shimada, Yusuke and Wada, Takeshi and Kim, Hyoung Seop and Takeuchi, Akira and Konno, Toyohiko J and Kato, Hidemi and Okulov, Ilya V},
  journal={Advanced Materials},
  volume={32},
  number={6},
  pages={1906160},
  year={2020},
  publisher={Wiley Online Library}
}

@article{siddique2023diamond,
  title={Diamond-structured nanonetwork gold as mechanical metamaterials from bottom-up approach},
  author={Siddique, Suhail K and Sadek, Hassan and Wang, Chi-Wei and Lee, Chang-Chun and Tsai, Cheng-Yuan and Chang, Shou-Yi and Li, Chia-Lin and Hsueh, Chun-Hway and Ho, Rong-Ming},
  journal={NPG Asia Materials},
  volume={15},
  number={1},
  pages={36},
  year={2023},
  publisher={Springer Japan Tokyo}
}

@article{zhao2020open,
  title={Open-cell tungsten nanofoams: Chloride ion induced structure modification and mechanical behavior},
  author={Zhao, Mingyue and Pfeifenberger, Manuel J and Kiener, Daniel},
  journal={Results in Physics},
  volume={17},
  pages={103062},
  year={2020},
  publisher={Elsevier}
}

@article{baker2024optimal,
  title={Optimal indent spacing for instrumented nanoindentation of nanoporous gold},
  author={Baker, Kerry A and Balk, T John},
  journal={MRS Communications},
  volume={14},
  number={1},
  pages={90--95},
  year={2024},
  publisher={Springer}
}

@article{mccue2018gaining,
  title={Gaining new insights into nanoporous gold by mining and analysis of published images},
  author={McCue, Ian and Stuckner, Joshua and Murayama, Mitsu and Demkowicz, Michael J},
  journal={Scientific reports},
  volume={8},
  number={1},
  pages={6761},
  year={2018},
  publisher={Nature Publishing Group UK London}
}

@article{kwon2021compressive,
  title={Compressive properties of nanoporous gold through nanoindentation: an analytical approach based on the expanding cavity model},
  author={Kwon, Oh Min and Kim, Jiyeon and Lee, Jinwoo and Kim, Jong-hyoung and Ahn, Hee-Jun and Kim, Ju-Young and Kim, Young-Cheon and Kwon, Dongil},
  journal={Metals and Materials International},
  volume={27},
  number={10},
  pages={3787--3795},
  year={2021},
  publisher={Springer}
}

@article{champion2019understanding,
  title={Understanding the interdependence of penetration depth and deformation on nanoindentation of nanoporous silver},
  author={Champion, Yannick and Laurent-Brocq, Mathilde and Lhuissier, Pierre and Charlot, Fr{\'e}d{\'e}ric and Moreira Jorge Junior, Alberto and Barsuk, Daria},
  journal={Metals},
  volume={9},
  number={12},
  pages={1346},
  year={2019},
  publisher={MDPI}
}

@article{cheng2023investigation,
  title={Investigation of the Influence of Nanoscale Porosity in the Interfacial Layers on the Mechanical Properties of Helium Plasma-Exposed Tungsten},
  author={Cheng, Li and Patino, Marlene and Chambers, Robert J and Cai, Shengqiang and Baldwin, Matthew and Bandaru, Prabhakar},
  journal={ACS Applied Engineering Materials},
  volume={1},
  number={7},
  pages={1822--1830},
  year={2023},
  publisher={ACS Publications}
}

@article{lee2025development,
  title={Development of 3D interconnected nanoporous TiZrHfNbTaNi high-entropy alloy via liquid metal dealloying and subsequent synthesis of (TiZrHfNbTaNi) O high-entropy oxide},
  author={Lee, Jae Hyuk and Ha, Soo Vin and Seong, Jihye and Takeuchi, Akira and Song, Ruirui and Kato, Hidemi and Joo, Soo-Hyun},
  journal={Journal of Materials Research and Technology},
  volume={35},
  pages={5204--5215},
  year={2025},
  publisher={Elsevier}
}

@article{lee2026nanoporous,
  title={Nanoporous high-entropy alloys fabricated by liquid metal dealloying and surface anodizing for enhanced electrochemical performance in capacitors},
  author={Lee, Jae Hyuk and Seong, Jihye and Joo, Soo-Hyun and Kato, Hidemi},
  journal={Journal of Materials Research and Technology},
  year={2026},
  publisher={Elsevier}
}

@article{choi2026formation,
  title={Formation of 3D interconnected multiphase heterostructures from CoCrFeMnNi high-entropy alloy via liquid metal dealloying using Cu--Ag melts},
  author={Choi, Munsu and Cho, Hogeun and Lee, Jae Hyuk and Takeuchi, Akira and Kato, Hidemi and Kim, Hyoung Seop and Hong, Soon-Jik and Choi, Yongseok and Han, Seung Zeon and Joo, Soo--Hyun},
  journal={Journal of Materials Research and Technology},
  year={2026},
  publisher={Elsevier}
}

@article{zou2024ligament,
  title={Ligament morphology and elastic modulus of porous structure formed by liquid metal dealloying},
  author={Zou, Lijie and Shao, Jun-Chao and Jin, Hai-Jun},
  journal={Journal of Materials Research and Technology},
  volume={31},
  pages={3914--3920},
  year={2024},
  publisher={Elsevier}
}

@article{chen2012structural,
  title={Structural evolution of nanoporous gold during thermal coarsening},
  author={Chen-Wiegart, Yu-chen Karen and Wang, Steve and Chu, Yong S and Liu, Wenjun and McNulty, Ian and Voorhees, Peter W and Dunand, David C},
  journal={Acta Materialia},
  volume={60},
  number={12},
  pages={4972--4981},
  year={2012},
  publisher={Elsevier}
}

@article{melnikova2021nanomechanical,
  title={Nanomechanical and nanotribological properties of nanostructured coatings of tantalum and its compounds on steel substrates},
  author={Melnikova, Galina and Kuznetsova, Tatyana and Lapitskaya, Vasilina and Petrovskaya, Agata and Chizhik, Sergei and Zykova, Anna and Safonov, Vladimir and Aizikovich, Sergei and Sadyrin, Evgeniy and Sun, Weifu and others},
  journal={Nanomaterials},
  volume={11},
  number={9},
  pages={2407},
  year={2021},
  publisher={MDPI}
}

@article{khanuja2009xps,
  title={XPS depth-profile of the suboxide distribution at the native oxide/Ta interface},
  author={Khanuja, Manika and Sharma, Himani and Mehta, BR and Shivaprasad, SM},
  journal={Journal of Electron Spectroscopy and Related Phenomena},
  volume={169},
  number={1},
  pages={41--45},
  year={2009},
  publisher={Elsevier}
}

@article{erlebacher2003pattern,
  title={Pattern formation during dealloying},
  author={Erlebacher, J and Sieradzki, Karl},
  journal={Scripta materialia},
  volume={49},
  number={10},
  pages={991--996},
  year={2003},
  publisher={Elsevier}
}

@article{wu2025formation,
  title={Formation of nanoporous Ta structure on Ti-Ta alloy based on a novel surface dealloying process},
  author={Wu, Lunemin and Xv, Qihang and Cao, Sheng and Zhou, Jian and You, Deqiang and Wang, Xiaojian},
  journal={Journal of Alloys and Compounds},
  volume={1022},
  pages={179945},
  year={2025},
  publisher={Elsevier}
}

@article{song2020liquid,
  title={Liquid metal dealloying of titanium-tantalum (Ti-Ta) alloy to fabricate ultrafine Ta ligament structures: A comparative study in molten copper (Cu) and Cu-based alloys},
  author={Song, Tingting and Tang, HP and Li, Yaning and Qian, Ma},
  journal={Corrosion Science},
  volume={169},
  pages={108600},
  year={2020},
  publisher={Elsevier}
}

\end{document}